\documentclass[usegraphicx,usenatbib]{mn2e}

\usepackage{amssymb}

\begin{document}

\title[What is the best way to measure BAO?]
{
What is the best way to measure baryonic acoustic oscillations?
}
\author[Ariel G. S\'{a}nchez et al.]
{\parbox[t]{\textwidth}{
Ariel G. S\'{a}nchez$^{1,2}$\thanks{E-mail: arielsan@mpe.mpg.de},
C. M. Baugh$^{3}$,
R. Angulo$^{3}$.}
\vspace*{6pt} \\ 
$^{1}$ Instituto de Astronom\'ia Te\'orica y Experimental - Observatorio Astron\'omico C\'ordoba, Laprida 854, X5000BGR, C\'ordoba, Argentina.\\
$^{2}$ Max-Planck-Institut f\"ur Extraterrestrische Physik, Giessenbachstrasse, 85748 Garching, Germany.\\
$^{3}$ The Institute for Computational Cosmology, Department of Physics, University of Durham, South Road, Durham DH1
3LE, UK.\\
}
\date{Submitted to MNRAS}
\maketitle

\begin{abstract}
Oscillations in the baryon-photon fluid prior to recombination 
imprint different signatures on the power spectrum and correlation 
function of matter fluctuations. The measurement of these features 
using galaxy surveys has been proposed as means to determine the 
equation of state of the dark energy. The accuracy required to 
achieve competitive constraints demands an extremely good understanding 
of systematic effects which change the baryonic acoustic oscillation 
(BAO) imprint. We use 50 very large volume N-body simulations to 
investigate the BAO signature in the two-point correlation function. 
The location of the BAO bump does not correspond to the sound horizon 
scale at the level of accuracy required by future measurements, even 
before any dynamical or statistical effects are considered. Careful 
modelling of the correlation function is therefore required to extract 
the cosmological information encoded on large scales. We find that the 
correlation function is less affected by scale dependent effects than 
the power spectrum.  We show that a model for the correlation function proposed 
by Crocce \& Scoccimarro (2008), based on renormalised perturbation theory, 
gives an essentially unbiased measurement of the dark energy equation of state.  
This means that information from the large scale shape of the correlation 
function, in addition to the form of the BAO peak, can be used to provide 
robust constraints on cosmological parameters. The correlation function 
therefore provides a better constraint on the distance scale 
($\sim 50\%$ smaller errors with no systematic bias) than the more 
conservative approach required when using the power spectrum (i.e. which requires 
amplitude and long wavelength shape information to be discarded).
\end{abstract}

\begin{keywords}
methods: N-body simulations – cosmology: theory – large-scale structure of
Universe.
\end{keywords}

\section{Introduction}

Prior to recombination, the ionised plasma of electrons and protons 
was coupled to the radiation in the Universe through the electromagnetic 
interaction. The pressure exerted by the photons worked against the  
gravitational collapse of perturbations in the density of baryons. 
This led to oscillations in the photon-baryon fluid. These ripples 
are imprinted on the temperature power spectrum of the cosmic microwave 
background ({\tt CMB}), as detected convincingly for the first time 
around the turn of the millennium \citep{debernardis00,hanany00}.
A series of acoustic peaks has now been measured with impressive precision by the {\tt WMAP} satellite 
\citep{bennet03,hinshaw03,hinshaw06,hinshaw08} and the Arcminute Cosmology Bolometer Array 
Receiver \citep[{\tt ACBAR},][]{reichardt08}. The positions and relative heights of these peaks can be used to place tight constraints on the values of the fundamental cosmological parameters, particularly
when the {\tt CMB} measurements are combined with measurements of the
galaxy power spectrum \citep{efstathiou02,percival02,spergel03,tegmark04,seljak05,sanchez06, seljak06, spergel07}.

The acoustic oscillations are also imprinted on the matter power spectrum, 
albeit with different phases from the features seen in the CMB spectrum 
and with a reduced amplitude, due to the small fraction of the total mass 
in the Universe believed to be in the form of baryons \citep{sugiyama95,EH98,meiksin1999}. 
Recently, these features have attracted a great 
deal of interest as a potential route to measuring the equation of state 
of the dark energy, $w_{\rm DE}= P_{\rm DE}/\rho_{\rm DE}$, where $P_{\rm DE}$
is the pressure of the dark energy 
and $\rho_{\rm DE}$ is its density 
\citep{blake03,hu03,linder03,seo03,wang06,guzik2007,seo2007,seo2008}. The acoustic oscillations in the 
matter power spectrum are related to the sound horizon scale. If we know 
the size of the sound horizon through measurements of the CMB temperature 
fluctuations, we can use this scale as a standard ruler. The apparent size 
of this ruler depends upon the parameter $w_{\rm DE}$, as this influences the 
angular diameter distance out to a given redshift.

In this paper, we focus our attention on the two-point correlation function 
which is the Fourier transform of the power spectrum. The series of
acoustic oscillations seen in the power spectrum translates into a bump
or broad spike in the correlation function \citep{matsubara04}. Such a feature was detected
for the first time in the correlation function of luminous red galaxies
\citep{eisenstein05,okumura08,estrada08}. Evidence has also been found for
baryon oscillations in the power spectrum of galaxy clustering
\citep{cole05,hutsi06,padmanabhan06,percival07b}. The appearance of the 
acoustic oscillations in the power spectrum was considered in 
an earlier companion paper to this one by \citet{angulo08}.

Whilst the bulk of the literature on acoustic oscillations has focused on the power spectrum,
there has been renewed interest recently in the use of the correlation function to extract the
sound horizon scale. \citet{angulo05} showed that the correlation function of a sample of clusters
at $z \sim 1$, as anticipated from the Sunyaev-Zeldovich survey proposed with the South Pole Telescope \citep{SPT} 
or a red sequence photometric survey with VISTA (http://www.vista.ac.uk), could potentially constrain
the sound horizon scale to an accuracy of $\sim2\%$, in turn fixing the dark energy equation of state
to better than $10\%$.

For the acoustic oscillation approach to provide competitive estimates of the value of $w_{\rm DE}$, one
needs to extract the sound horizon scale from the measured correlation function to sub-percent accuracy. To achieve this goal it is essential to quantify any
systematic deviations of the scale recovered from the size of the sound horizon. In order to do so in a way
that avoids the introduction of new biases or systematic errors, a complete understanding of all the processes
that shape the observed correlation function is required.

Based on the fitting formula of \citet{smith2003} and using the position of the acoustic peak in the correlation
function as an estimator of the sound horizon at recombination, \citet{guzik2007} concluded that non-linear evolution biases the constraints by less than $0.3\%$. \citet{eisenstein06b} argue that any possible shift of the acoustic scale
is unobservably small even at $z=0$. On the other hand, the more recent analyses of \citet{smith08} and \citet{crocce08} have shown that both large volume numerical simulations and theoretical predictions based on renormalized perturbation theory indicate that the position of the peak in the correlation function shows important shifts with respect to the linear theory prediction. If unaccounted for, these shifts bias the constraints obtained by using BAO measurements as a standard ruler.

In this paper we analyse some of the different tests that have been proposed to
extract cosmological information from the correlation function on large scales,
with particular emphasis on the possible systematic errors that can be introduced
in their application.
We first explain the relationship between the size of the sound horizon and the
location and form of the peak in the correlation function in \S~\ref{sec:peak_pos}.
We next determine the accuracy of various means of computing the matter two-point
correlation function according to linear perturbation theory for a fixed set of cosmological
parameters (\S~\ref{sec:full_shape}).

Besides the linear evolution of density perturbations, there are other processes
that will affect a measurement of the correlation function obtained from the new large galaxy
redshift surveys: non-linear evolution of the density
field, redshift-space distortions and halo bias. It is extremely important to
test our ability to model these processes in order to firmly assess if the
precision claimed by some future experiments is attainable with our current
understanding of these effects.
In \S~\ref{sec:nbody} we describe the ensemble of simulations we used to model these effects, and the methodology implemented to estimate the two-point correlation function. In \S~\ref{sec:modelling_full_shape} we describe the details of our modelling of the correlation function and its ability to reproduce the results from the N-body simulations with respect to non-linear evolution, redshift-space distortions and scale-dependent halo bias.
The measurement of the imprints of the baryonic acoustic oscillations in the power spectrum and correlation functions are affected in different ways by these problems. In \S~\ref{sec:pk_vs_xi} we compare the performace of these 
statistics to see which one offers the most advantages as a tool to recover unbiased constraints on cosmological parameters from BAO measurements.
Finally, in \S~\ref{sec:conclusions} we give a summary of our main results.

\section{What is the relation between the sound horizon scale and the peak in the 
correlation function?}
\label{sec:peak_pos}

The two-point correlation function has a bump on scales in excess of $100 h^{-1}{\rm Mpc}$ \citep{eisenstein05}. 
It is often assumed that the position of this bump coincides {\it exactly} with the 
sound horizon scale.
In this section, we explore the connection between the sound horizon, 
the nature of the acoustic oscillations imprinted on the matter power spectrum and the position and form of the 
acoustic bump in the correlation function. We shall demonstrate that, at the level of accuracy required
by forthcoming measurements of distances using BAO, the assumption that the sound horizon scale is equal to the 
position of the peak in the correlation function is incorrect and introduces a systematic error in the distance 
measurement which is in excess of the expected random errors. 

\begin{figure}
\centering
\centerline{\includegraphics[width=0.5\textwidth]{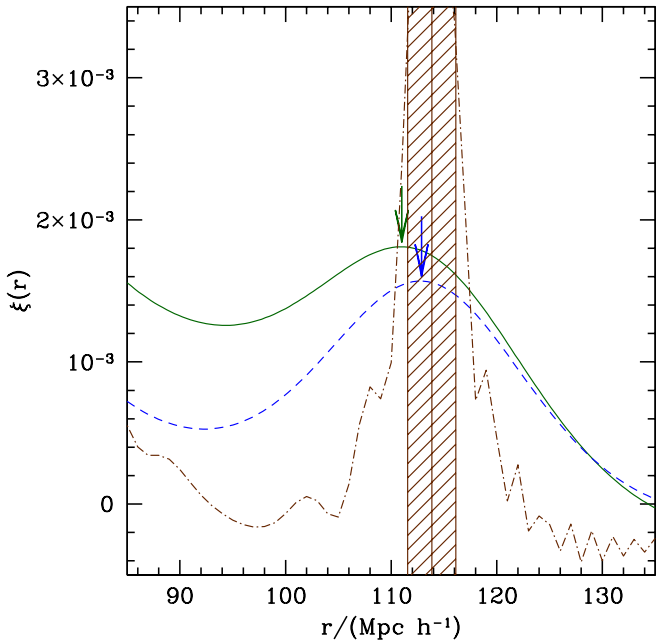}}
\caption{
The correlation functions computed by Fourier transforming the power spectrum 
obtained using the EH98 formula for $P(k)$ for three cases:  solid line - a 
fully consistent linear theory $P(k)$, dot-dashed line - a $P(k)$ with no Silk 
damping and with dominant velocity overshoot on all wavenumbers and dashed - 
a $P(k)$ with no Silk damping but a standard velocity overshoot. 
The arrows mark the location of the peak in the acoustic bump, defined as the 
local maximum. The shaded region indicates a 2\% error on the sound 
horizon scale, which is shown by the central solid line. 
}
\label{fig:zoom_xi}
\end{figure}

The correlation function is the Fourier transform of the power spectrum (see Eq.~(\ref{eq:xi}) in 
the next section). The acoustic 
oscillations of the baryon-photon fluid are imprinted on the matter power spectrum as a 
series of waves. {\it If} it was the case that these oscillations had a fixed 
wavelength and amplitude, then there would be a sharp feature in the correlation 
function at a scale centred on the wavelength of oscillation. However, it turns out, for 
various physical reasons, that the acoustic oscillations in the power spectrum have {\it neither} 
a fixed wavelength nor a fixed amplitude. This leads to a more complicated relationship  
between the sound horizon scale and the position of the peak in the correlation function 
than the naive assumption of equality. Furthermore, the size of the discrepancy between 
the peak location and the sound horizon scale depends upon the values adopted for the 
cosmological parameters, such as the matter density. 

The physical reasons behind the appearance of the acoustic oscillations are set out 
clearly by \citet[hereafter EH98]{EH98}, who discuss the form of the power spectrum of density
fluctuations in a universe containing cold dark matter and baryons. There are two phenomena 
which have a direct impact on the nature of the acoustic oscillations. The easiest to understand 
is the reduction in the amplitude of the oscillations with increasing wavenumber. This is due to 
an imperfect coupling between photons and baryons as the end of recombination is approached; photons 
diffuse out of perturbations and the remaining coupling with the baryons causes a Compton drag which 
leads to some baryons being pulled out of perturbations. This leads to a reduction in the amplitude 
of the density fluctuation known as Silk damping \citep{silk1968}. The explanation of the change 
in the wavelength of the oscillation with harmonic, at wavenumbers for which $ks \le 10 $, where $s$ 
is the sound horizon, is more subtle. After recombination, the baryons can `slip' past the photons. 
This motion of the baryons, called the velocity overshoot, generates new perturbations 
\citep{SZ1970,PV1980}. During recombination, EH98 argue that the velocity overshoot is not 
the dominant contribution to the growth of the baryonic perturbation, which leads to 
the ``node shift'' in the baryonic transfer function. 

The physically motivated fitting formula for the matter power spectrum presented by EH98 
can be used to show the impact of the phenomena described above on the form of the acoustic
bump in the correlation function. The consequences of Silk damping and the impact of velocity 
overshoot can be readily identified in this formula and can therefore be switched on or off 
to investigate their impact on the correlation function. 
We consider three cases: 
\begin{itemize}
\item[(1)] The fully consistent power spectrum, with Silk damping and a consistent treatment of velocity overshoot.
\item[(2)] A toy model in which there is no Silk damping and velocity overshoot is 
dominant at all $k$. This corresponds to setting $k_{\rm silk} \rightarrow \infty$ 
and $\bar{s}=s$ in eqn.~21 of EH98 ($\bar{s}$ is defined in eqn.~22 of EH98). 
\item[(3)] A second toy model with Silk damping (the appropriate value of $k_{\rm silk}$ is used) but with 
velocity overshoot dominant for all $k$ (again, retaining $\bar{s}=s$ in eqn.~21 of EH98).
\end{itemize}
We then Fourier transform the resulting power spectrum in each case to obtain the correlation function 
of matter fluctuations in linear perturbation theory.

The results of considering the above test cases are presented in Fig.~\ref{fig:zoom_xi}, where we assume the 
background cosmological parameters to be $\Omega_{0}=0.25, \Omega_{\rm b}=0.041$ and $h=0.7$, 
where $H_{0}=100 h {\rm km s}^{-1}{\rm Mpc}^{-1}$. The sound horizon scale for these 
cosmological parameters is $s=113.8 h^{-1}$Mpc (using Eq. 6 of EH98), and is marked by the 
vertical line in Fig.~\ref{fig:zoom_xi}. Using the full linear theory $P(k)$ (Case 1 above) to compute 
$\xi(r)$ yields a broad peak with a maximum at $r=111.0 h^{-1}{\rm Mpc}$. The correlation function 
corresponding to Case 2 has a much sharper bump with a higher amplitude than in Case 1. The maximum of 
this feature is at the true sound horizon scale. Turning on Silk damping (Case 3) shifts the peak 
to $ r = 113.0 h^{-1}{\rm Mpc}$. The shaded area around the vertical line in Fig.~\ref{fig:zoom_xi} 
represents a 2\% error on the sound horizon measurement. This is size of the random error forecast for the ongoing 
WiggleZ experiment \citep{glazebrook2007,angulo08}. The peak of the correlation function bump 
lies outside the shaded region. Therefore approaches which advocate measuring the location of the bump in the 
correlation function will make significant systematic errors if this scale is mistakenly identified as the 
sound horizon scale.

\begin{figure}
\includegraphics[width=0.47\textwidth]{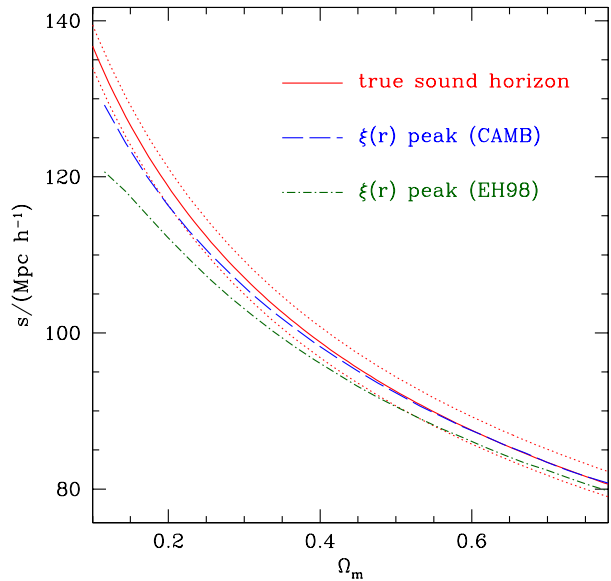}
\caption{
The true sound horizon, $s$ (solid line), and the position of the acoustic peak 
in the correlation function, $r_p$ (dot-dashed line), as a function of $\Omega_{\rm m}$ 
for a fixed value of $\Omega_{\rm b}=0.041$. The dotted lines indicate a 2\% spread in 
the value of the sound horizon. The dashed line shows location of the acoustic bump when 
the EH98 formalism for $P(k)$ is replaced by a more accurate calculation made with CAMB 
(see text for details). 
}
\label{fig:peak_pos}
\end{figure}

The size of the discrepancy between the location of the acoustic bump in the correlation function and 
the sound horizon scale depends upon the cosmology. Fig.~\ref{fig:peak_pos} shows the variation of the sound 
horizon, $s$ (solid line), and the acoustic bump location, $r_{\rm p}$ (dot-dashed line), as a function of $\Omega_{\rm m}$ 
for a fixed value of $\Omega_{\rm b}=0.041$. It can clearly be seen that, although these two scales agree for
$\Omega_{\rm m} > 0.4$, the scale $r_{\rm p}$ is smaller than $s$ for values of the density parameter that are 
consistent with current observational constraints \citep[e.g.][]{sanchez06}. The dotted lines in Fig.~\ref{fig:peak_pos} 
indicate a 2\% error on $s$. For smaller values of $\Omega_{\rm m}$, where the baryon fraction is high and then the effects 
described above more important, the deviation between $s$ and $r_p$ can be larger than this 2\% limit.
Therefore, a simple association of the position of the peak with the true sound horizon will lead to results 
that are biased towards high values of $\Omega_m$. This highlights the importance of the correct interpretation 
of the observations in order to avoid the introduction of biases and systematic errors when using BAO observations 
as cosmological probes.

In the next section we will see that there are differences between the correlation functions computed using the EH98 fitting 
formula and that obtained using CMBFAST \citep{cmbfast} and CAMB \citep{camb}. The value of $r_{p}$ inferred using the EH98 
approximation for $P(k)$ is biased towards smaller scales (see the dot-dashed line in Fig.~\ref{sec:peak_pos}). Nevertheless, 
the qualitative description of the processes that shift the location of the bump in $\xi(r)$ peak away from $s$ is still correct. 

The conclusion from this simple analysis is that, even when using linear perturbation theory, it is wrong to assume that
the position of the peak in the correlation function corresponds exactly with the value of the sound horizon. Such an assumption will lead to biased constraints on cosmological parameters. The correct approach is to model the full shape of the correlation function which we pursue in the following sections.

\section{The linear theory two-point correlation function}
\label{sec:full_shape}

In Section 2, we demonstrated that it is not possible to use the position of the acoustic peak in the 
correlation function as a proxy for the sound horizon scale. In order to obtain useful constraints on 
the cosmological parameters from the correlation function, it will be necessary to model the acoustic bump. 
In this Section, we consider the form of the matter correlation function on large scales in linear 
perturbation theory. For a realistic survey, the impact on the correlation function of the non-linear 
growth of perturbations, redshift space distortions and bias will have to be taken into account. We 
introduce measurements of the correlation function from N-body simulations in Section 4 and then 
discuss a full model of the correlation function in Section 5. 

The first step to a model of $\xi(r)$ is the reproduction of the matter correlation function in 
linear perturbation theory. We obtain $\xi(r)$ by Fourier transforming the linear theory mass power spectrum
\begin{equation}
\xi(r) = \int_{-\infty}^{\infty} \Delta^{2}(k) j_0(kr){\rm d}\ln(k), \label{eq:xi}
\end{equation}
where $\Delta^{2}(k)=P(k)k^3/(2\pi^2)$ is the dimensionless power spectrum and $j_0(y)$ is the spherical Bessel function.

\begin{figure}
\includegraphics[width=0.47\textwidth]{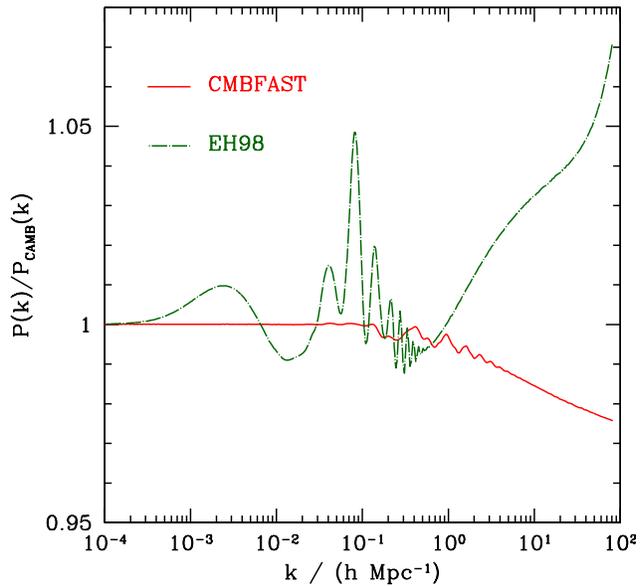}
\caption{
The ratio of the power spectra obtained using the EH98 fitting formula (dot-dashed line) 
and CMBFAST (solid line) to the one obtained using CAMB. 
}
\label{fig:pk_ratio}
\end{figure}

An accurate correlation function in linear perturbation theory therefore requires an accurate calculation of the 
power spectrum. The most commonly used codes to compute the power spectrum are: i) the approximate formula of EH98, 
ii) CMBFAST \citep{cmbfast} and iii) CAMB \citep{camb}. \citet{seljak2003} tested the accuracy of CMBFAST against 
full Boltzmann codes and found that the relative error in the matter power spectrum is below $10^{-3}$. This comparison 
did not include CAMB although a similar accuracy is claimed for this code. The last release of CMBFAST was in 2003; 
CAMB is under continual development and so we adopt the view that amongst these three alternatives, CAMB provides the 
benchmark. As we have already mentioned in Section 2, EH98 produced a physically motivated functional form for the 
power spectrum, which they calibrated against CMBFAST. EH98 found that their formalism reproduced the results of 
CMBFAST at the percent level, over a restricted range of wavenumbers. EH98 provided subroutines for their fitting 
formulae, which are now in common use and are faster than running CMBFAST or CAMB.

Fig.~\ref{fig:pk_ratio} shows the ratio of the power spectra obtained using the EH98 fitting formula (dot-dashed line) 
and CMBFAST (solid line) to the spectrum obtained using CAMB for a flat cosmological model with $\Omega_{\mathrm m}=0.237, 
\Omega_{\rm b}=0.041$ and $h=0.735$ (which corresponds to the best fitting parameters from \citet{sanchez06}). 
There are clear differences between the $P(k)$ of EH98 and CAMB on the scales relevant to the acoustic oscillations 
that can be as large as 5\%. For this particular model, CAMB and CMBFAST show impressively good agreement up to 
$k \sim 1 h {\rm Mpc}^{-1}$.

\begin{figure}
\includegraphics[width=0.47\textwidth]{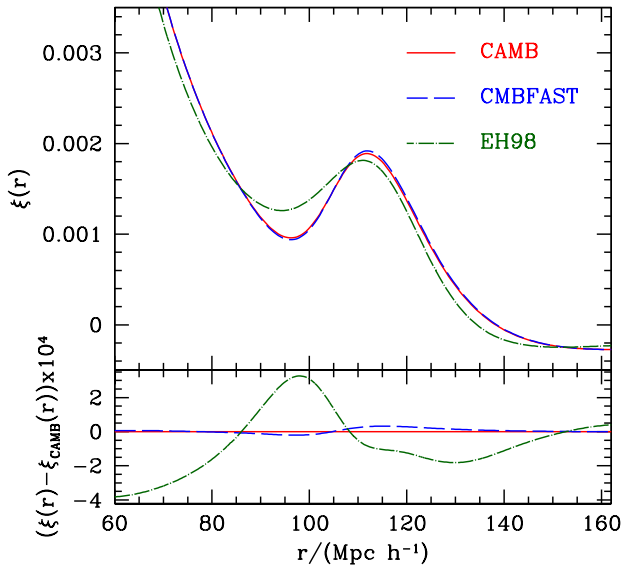}
\caption{
Upper panel: The correlation functions obtained by Fourier transforming the power
spectra computed using the EH98 fitting formula (dot-dashed line), CAMB (solid
line) and CMBFAST (dashed line). Lower panel: the residuals of the correlation
functions obtained from the EH98 and CMBFAST power spectra with respect to that 
obtained using CAMB.
}
\label{fig:xi_diff}
\end{figure}
The consequences for the correlation function of these differences in the predicted linear theory power spectrum can 
be seen in Fig.~\ref{fig:xi_diff}. The upper panel of Fig.~\ref{fig:xi_diff} shows the linear theory correlation functions 
obtained by Fourier transforming the power spectra shown in Fig.~\ref{fig:pk_ratio} according to Eq.~(\ref{eq:xi}). 
It can be clearly seen that while the correlation functions derived from the $P(k)$ computed with CAMB and 
CMBFAST are in very good agreement, the use of the approximate $P(k)$ of EH98 produces a correlation function with 
a quite different shape and peak position. The lower panel of Fig.~\ref{fig:xi_diff} shows the residuals of the correlation 
functions obtained from the EH98 and CMBFAST power spectra from the CAMB result. For the case of the EH98 formula, 
these residuals are comparable to or larger than the variance expected in future galaxy surveys (see \S~.\ref{sec:variance}). 
These results show that the use of EH98 formula to model the shape of $\xi(r)$ can introduce strong biases on the obtained 
constraints. 
Our findings are consistent with those of \citet{sanchez08} who showed that the use of the EH98 
fitting formula to model the shape of the galaxy power spectrum introduces changes in the recovered value 
of $\Omega_{\mathrm m} h$ of the order of one sigma. The EH98 formalism is excellent for providing physical insight into 
the form of the power spectrum in a cold dark matter universe and is the simplest and quickest of the prescriptions listed above 
to use to generate large numbers of model spectra. However, it was never intended to supercede the more accurate calculations 
from codes like CMBFAST. In order to attain the level of precision demanded by forthcoming BAO analyses, the EH98 formalism 
should be replaced by the more accurate calculation of CAMB when modelling power spectra or correlation functions.

\section{Numerical modelling of the correlation function}
\label{sec:nbody}

A number of effects can alter the form of the two-point correlation function from  
the linear perturbation theory predictions presented in the last section: the non-linear 
growth of perturbations, redshift space distortions and bias \citep{guzik2007,smith07,angulo08,crocce08,smith08}. In this section 
we model the impact of these processes on the two-point correlation function using N-body simulations. 
The ensemble of simulations used are described in \S~\ref{sec:ensemble}. The estimation of the correlation function 
in a large number of huge volume simulations could be prohibitively expensive without an efficient 
algorithm which we describe in \S~\ref{sec:estimator}. Finally, the use of an ensemble of simulations allows a direct 
estimate of the errors on the measured correlation function, which we set out in \S~\ref{sec:variance}.

\subsection{The ensemble of N-body simulations}
\label{sec:ensemble}

We use an ensemble of 50 moderate resolution, large volume N-body simulations called 
{\tt L-BASICC}\,{\tt II} \citep{angulo08}. These simulations are analogous to the {\tt L-BASICC} 
ensemble of simulations employed by Angulo et~al. to assess the detectability of the acoustic 
oscillations in power spectrum measurements from future galaxy surveys. The only difference is that  
the {\tt L-BASICC}\,{\tt II} runs are based on a different choice of the cosmological parameters.
We adopt a $\Lambda$CDM cosmology consistent with current constraints from 
the cosmic microwave background data and large scale structure measurements \citep{sanchez06, spergel07}. 
We assume a flat $\Lambda$CDM cosmological model with matter density parameter, $\Omega_{\rm m}=0.237$, 
baryonic density parameter, $\Omega_{\rm b}=0.041$, scalar spectral index, $n_{\rm s} = 0.954$, a 
normalisation of density fluctuations, $\sigma_{8} = 0.77$, and Hubble constant, $h=0.735$.

Each of the {\tt L-BASICC}\,{\tt II} simulations covers a comoving cubical region of side 
$1340\,h^{-1}\,$Mpc, with the dark matter followed using $448^3$ particles. This gives 
a particle mass comparable to that employed in the Hubble Volume simulation \citet{evrard2002}.
The equivalent Plummer softening length in the gravitational force is $\epsilon = 200\,h^{-1}\,\rm{kpc}$. 
The volume of each computational box, $2.41\,h^{-3}\,{\rm Gpc}^{3}$, is almost twenty times 
that of the Millennium Simulation \citep{springel2005}, and more than three times the volume of the
catalogue of luminous red galaxies from the SDSS used to make the first detection of the acoustic
peak by \citet{eisenstein05}. The total volume of the ensemble is $120\,h^{-3}\,{\rm Gpc}^{3}$, 
more than four times that of the Hubble Volume. The position and velocity of every particle are stored 
at 3 output times (z = 0.0, 0.5, 1.0).  We produce a friends-of-friends halo catalogue at each redshift 
of objects with ten or more particles (corresponding to a mass limit of $1.75 \times 10^{13}\,h^{-1}\,\rm{M_{\sun}}$). 
Due to their limited mass resolution, it is not feasible to populate these simulations with galaxies 
using semi-analytic models. 

The initial conditions for the simulations were generated by perturbing particles from a glass-like distribution \citep{baugh1995}. The input power spectrum of density fluctuations in linear perturbation theory is calculated using 
the {\tt CAMB} package of \citet{camb}. A different random seed is used for each member of the ensemble. The 
simulations were started at a redshift of $z=63$. \citet{angulo08} have shown that for this choice of the starting 
redshift, the scales relevant for the analysis of acoustic oscillations are unaffected by any transients introduced by the method used to generate the initial conditions. 

\subsection{The practical estimation of the two-point correlation function}
\label{sec:estimator}

The number of operations required to calculate the correlation function of $N_{\rm p}$ particles 
by direct pair counting scales as $N_{\rm p}^2$. This is infeasible for the large number of 
particles in our simulations on the large scales considered in our analysis. The situation is 
further exacerbated by the fact that we need to repeat the calculation many times, estimating 
the correlation function for three outputs in 50 simulations.

\begin{figure}
\includegraphics[width=0.47\textwidth]{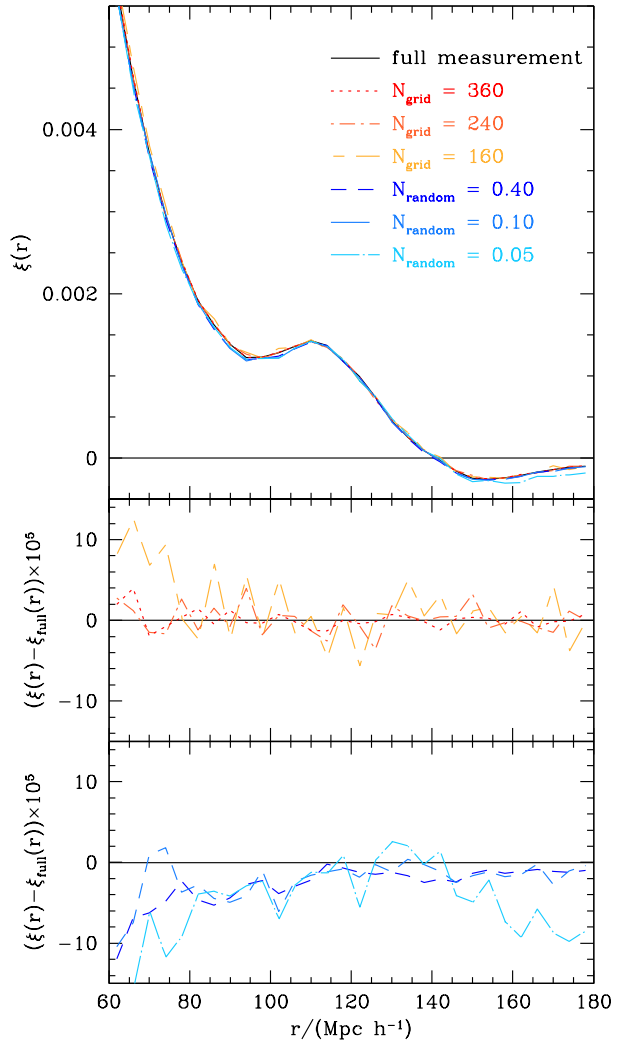}
\caption{
Comparison of the correlation function of one realization in our ensemble 
estimated by direct pair counting (solid line) with the ones obtained by 
counts of cubical cells with grid sizes of $N_{\rm grid}=160$ (short-long 
dashed line), $N_{\rm grid}=240$ (dot-dashed line) and $N_{\rm grid}=360$ 
(dotted line) and the ones obtained using a dilute sample containing a 
fraction of 5\% (dot-long dashed line), 20\% (long dashed line) and 40\% 
(dashed line) of the total sample (panel a). The deviations from the full (direct 
pair count) estimate can be better appreciated in panels (b) and (c), which 
show the difference $\xi_{\rm grid}(r)-\xi_{\rm full}(r)$ on a much expanded 
scale for the grid and for estimates derived from random samples of the 
dark matter particles respectively.
}
\label{fig:resol}
\end{figure}

\begin{figure*}
\centering
\centerline{\includegraphics[width=0.9\textwidth]{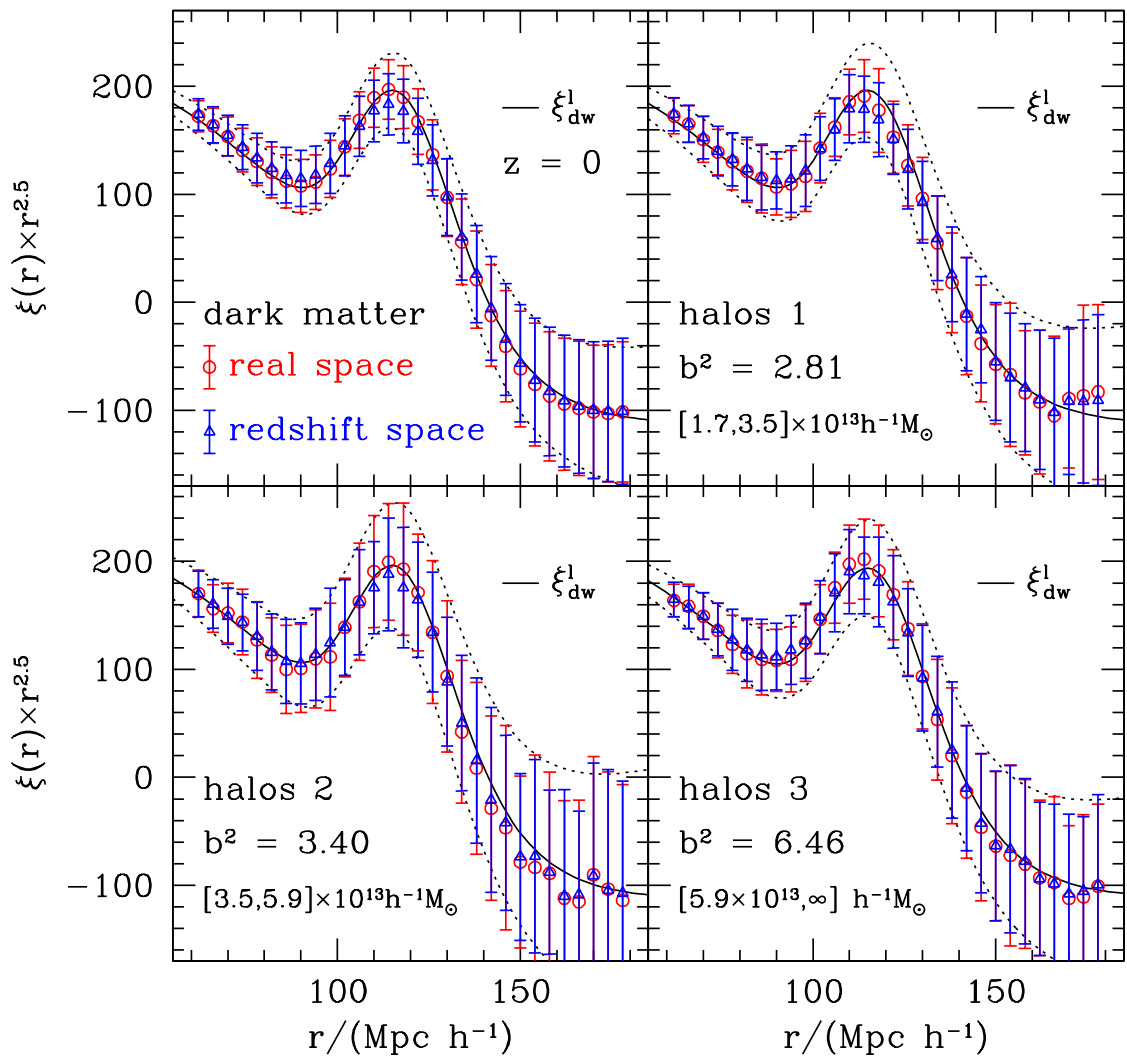}}
\caption{
Mean correlation functions at $z=0$ from our ensemble of simulations in real-space (circles) and redshift-space (triangles) for dark matter (panel (a)) and halos in samples 1, 2 and 3 (panels (b), (c) and (d)). The error bars show the variance from the estimates in the different realizations. To highlight the acoustic peak we show $\xi(r)\times r^{2.5}$. The results in redshift space are divided by the \citet{kaiser87} boost factor 
(see section~\ref{sec:redshift}).
The results for the halo correlation functions were scaled by the bias 
factors shown in the annotations in each panel (see section~\ref{sec:bias}). 
The solid lines show the fits to the simulation data with the model for 
the real space dark matter correlation function described in 
section~\ref{sec:non-linear}. The dotted lines in each panel show the 
estimates of the variance for each sample computed using Eq.~(\ref{eq:cov_xi2}). 
For the halo samples the variances are rescaled by the same bias factors 
than the measured correlation functions.
}
\label{fig:all_z0}
\end{figure*}

We follow the approach introduced by \citet{barriga2002} and \citet{eriksen2004} to speed up the 
estimation of the correlation function. The first step is to construct the density field of the 
simulation on a grid of $N_{\rm grid}$ cells using the nearest grid point (NGP) mass assignment scheme. 
Using this density grid, the correlation function can be estimated using 
\begin{equation}
 {\hat \xi}(r)=\frac{1}{N_{\rm pairs}}\sum_{i\,j}\delta_i\delta_j,
\end{equation}
where $\delta_i=(n_i-\left\langle n \right\rangle )/\left\langle n \right\rangle $ 
is the density fluctuation in the $i^{\rm th}$ bin of the grid and the sum extends 
over the $N_{\rm pairs}$ cells separated by distances between $r-\Delta r/2$ and 
$r-\Delta r/2$. This procedure scales as $N_{\rm grid}^2$ which is a big reduction 
in time since usually $N_{\rm grid} \ll N_p$. We use $N_{\rm grid}=240$. This method  
gives an accurate estimate of the correlation function on scales larger than a few 
grid cells. In this paper, we focus on pair separations in excess of $60 h^{-1}$Mpc, 
which corresponds to just over 10 cells. This approach could easily be adapted to 
work on smaller scales by using different size grids to tabulate the density field.

To further speed up the estimation of the correlation function, we first compute 
and store the indices of the $N_{\rm neigh}$ cells which contribute to a given bin 
of pair (cell) separation. This list of indices can then be translated to different 
locations on the density grid to find out which cells contribute to the estimate of 
the correlation function in each bin of pair separation.  In this way, no CPU time is 
wasted in re-computing cell separations. This reduces the number of operations from 
$N_{\rm grid}^2$ to $N_{\rm grid}N_{\rm neigh}$. A further speed up is possible as this 
algorithm can be naturally divided between processors and run in parallel.

\begin{figure*}
\centering
\centerline{\includegraphics[width=0.9\textwidth]{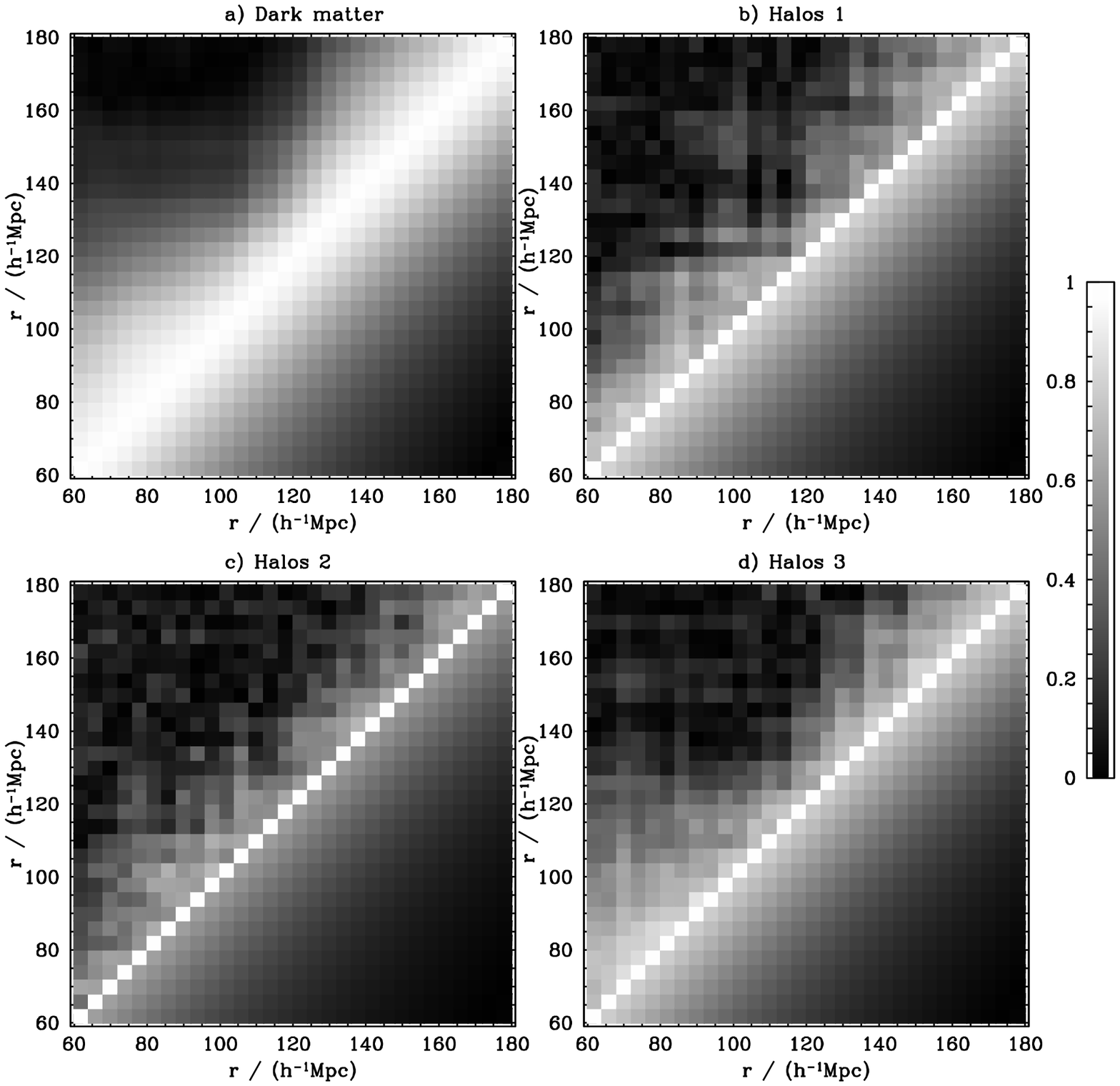}}
\caption{
Comparison of the correlation matrix $Cor(i,j)$ predicted by Eq.~(\ref{eq:cov_xi2}) (lower triangular parts of the matrices) and the ones obtained from the different realizations (upper triangular parts) for the dark matter (panel (a)), and the haloes in samples 1, 2 and 3 (panels (b), (c) and (d) respectively). In general, Eq.~(\ref{eq:cov_xi2}) describes accurately the elements close to the diagonal, particularly for the dark matter. In the more off-diagonal elements the estimation of $Cor(i,j)$ is noisy and the agreement is not so clear.
}
\label{fig:correl_mat}
\end{figure*}

The assignment of the particles to the grid represents a smoothing 
of the density field and may potentially have an impact on the 
shape of the estimated correlation function and hence could introduce 
a bias into our analysis. To assess the effect of this smoothing, 
we computed the correlation function of one realization in our 
ensemble of simulations using a direct pair count on all scales, which 
is computationally expensive. Fig.~\ref{fig:resol}(a) shows the comparison 
of this estimate (solid line) with those obtained from the counts of cubical 
cells with grid sizes of $N_{\rm grid}=160$ (short-long dashed line), 
$N_{\rm grid}=240$ (dot-dashed line) and $N_{\rm grid}=360$ (dotted line). 
The deviations from the direct pair count estimate, which we refer to as 
the ``full'' estimate from now on, are very small and can only be appreciated 
on a much expanded scale as plotted in Fig.~\ref{fig:resol}(b), which shows 
the difference $\xi_{\rm grid}(r)-\xi_{\rm full}(r)$ for the different values 
of $N_{\rm grid}$. It is clear that on using a value of $N_{\rm grid}=240$, 
the correlation function obtained gives an accurate description (at the level 
of a few parts in $10^5$) of $\xi_{\rm full}(r)$ over the range of scales 
relevant to our analysis. In particular, these deviations have no impact on 
the recovered values of $k_{\star}$ and $\alpha$ (see section 
\ref{sec:modelling_full_shape}).

An alternative approach to make a rapid estimate of the correlation function 
is to use a diluted sample of particles selected at random from the simulation. 
Fig.~\ref{fig:resol}(a) also shows the results obtained using a sample containing 
a fraction of 5\% (dot-long dashed line), 20\% (long dashed line) and 40\% 
(dashed line) of the total sample. The deviations with respect to the full 
estimate obtained using this approach are larger than in the previous cases in 
which a grid is used. This can be seen more clearly in Fig.~\ref{fig:resol}(c) 
which shows the difference $\xi_{\rm random}(r)-\xi_{\rm full}(r)$ for the 
different values of $N_{\rm ran}$. Even for a value of $N_{\rm ran} = 0.4$, 
the deviations from the full measurement are larger than in the grid based 
measurement, showing that the grid based approach is the preferred one 
when dealing with large samples where the full approach is not applicable. 
More importantly the correlation function estimated from the diluted samples 
displays a systematic distortion from $ \xi_{\rm full}$, as it is always 
below the full estimate and shows a different shape. This systematic deviation 
might introduce a small bias when constraining cosmological parameters.

We apply the algorithm described above to compute the two-point correlation function in real and 
redshift space for the three output redshifts ($z=0$, 0.5 and 1) for each of the 
{\tt L-BASIIC}\,{\tt II} simulations. Fig.~\ref{fig:all_z0}(a) shows the mean 
correlation function of the dark matter from the ensemble at $z=0$. The error bars 
show the variance in the correlation function estimated from the different realizations 
(see section~\ref{sec:variance}). To highlight the acoustic peak we plot $\xi(r)\times r^{2.5}$ 
instead of $\xi(r)$. The results in redshift space are divided by the \citet{kaiser87} 
``boost factor'' (see section~\ref{sec:redshift} for further details on the redshift space distortion). 
The solid lines in Figure~\ref{fig:all_z0} show the theoretical prediction for the real space 
dark matter correlation function computed as described in section~\ref{sec:non-linear}.

We have also measured the correlation function for halos in three mass bins at each redshift. 
Halo sample 1 includes haloes with masses $1.75\times10^{13} < (M/h^{-1}M_{\odot}) < 3.5\times10^{13}$, 
halo sample 2 covers the mass range $3.5\times10^{13} < (M/h^{-1}M_{\odot}) < 5.95\times10^{13}$, and 
halo sample 3 is all objects with mass in excess of $ M > 5.95\times10^{13} h^{-1} M_{\odot}$. 
At $z=0$, sample 1 accounts for roughly 50\% of all resolved haloes. The correlation functions 
measured for these halo samples are shown in Fig.~\ref{fig:all_z0}(b), (c) and (d). The results 
for the different samples have been scaled by the respective bias factors $b^2=2,78$, 3.48 and 6.38 
(for samples 1, 2 and 3) as described in the figure caption (see section~\ref{sec:bias} for further 
details on the distortions due to halo bias).

\subsection{The variance in $\xi(r)$}
\label{sec:variance}

In this section we compare the variance in the correlation function estimated from the 
ensemble of simulations with a theoretical prediction.  

The starting point in the analytical estimate is the assumption that the variance 
in the power spectrum is that expected for Gaussian fluctuations with a shot-noise 
component arising from the finite number of objects used to trace the density 
field \citep{FKP}:  
\begin{equation}
\sigma_{P}(k)=\sqrt{\frac{2}{V}}\left(P(k)+\frac{1}{{\bar n}} \right),
\label{eq:sigmap}
\end{equation}
where $V$ is the simulation volume, and ${\bar n}$ is the mean density of the objects 
considered (dark matter particles or halos). \citet{angulo08} found good agreement 
between this expression and the variance in $P(k)$ measured from numerical simulations. 

The covariance of the two-point correlation function is defined by \citep{cohn2006,smith08}:
\begin{eqnarray}
 C_{\xi}(r,r')&\equiv&\left\langle (\xi(r)-{\bar \xi}(r))(\xi(r')-{\bar \xi}(r'))\right\rangle \\
&=&\int \frac{{\rm d}k\,k^2}{2\pi^2}j_0(kr)j_0(kr')\sigma^2_P(k),
\label{eq:cov_xi}
\end{eqnarray}
where the last term can be replaced by Eq.~(\ref{eq:sigmap}).
The variance in the correlation function is simply $\sigma^2_{\xi}(r)=C_{\xi}(r,r)$. 
The direct application of Eq.~(\ref{eq:cov_xi}) would, however, lead to a substantial overprediction 
of the variance, since it ignores the effect of binning in pair separation which reduces the covariance 
in the measurement \citep{cohn2006,smith08}.

\begin{figure}
\includegraphics[width=0.47\textwidth]{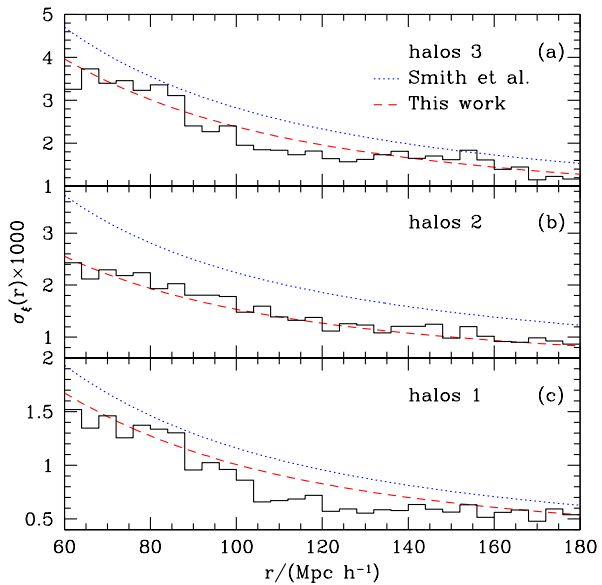}
\caption{
Comparison of the variance measured from our ensemble of simulations 
(solid line histogram) for the halo correlation functions in halo 
samples 1, 2 and 3 (lower panel, middle panel and upper panel respectively) 
with the predictions from Eqs.~(\ref{eq:cov_xi2}) (dashed lines) 
and~(\ref{eq:aprox_smith}) (doted lines).
}
\label{fig:variance}
\end{figure}

An estimate of the correlation function in the $i^{\rm th}$ pair separation bin 
${\hat \xi}_i$ corresponds to the shell averaged correlation function
\begin{equation}
 {\hat \xi}_i=\frac{1}{V_i}\int_{V_i} \xi(r)\,{\rm d}^3r,
\end{equation}
where $V_i$ is the volume of the shell. The covariance of this estimate is given by
\begin{eqnarray}
C_{{\hat \xi}}(i,j)&=&\frac{1}{V_i\,V_j}\int {\rm d}^3r\int {\rm d}^3r'C_{\xi}(r,r')\\
&=&\int \frac{{\rm d}k\,k^2}{2\pi^2}{\bar j}_0(k,i){\bar j}_0(k,j)\sigma^2_P(k),
\label{eq:cov_xi2}
\end{eqnarray}
where
\begin{equation}
{\bar j}_0(k,i)=\frac{1}{V_i}\int_{V_i} j_0(kr)\,{\rm d}^3r.
\end{equation}
The dotted lines in each panel of Fig.~\ref{fig:all_z0} show the estimates of the 
variance for the various samples computed using Eq.~(\ref{eq:cov_xi2}). 
For the halo samples the variance is rescaled by the same bias factors as 
the measured correlation functions. In all cases the variances measured from the 
ensemble and the theoretical predictions are in very good agreement. 

Fig.~\ref{fig:correl_mat} shows a comparison of the correlation matrix 
$Cor_{{\hat \xi}}(i,j)=C_{{\hat \xi}}(i,j)/\sigma_{{\hat \xi}}(i)\sigma_{{\hat \xi}}(j)$ 
predicted by Eq.~(\ref{eq:cov_xi2}) with the estimate from the ensemble of simulations. 
The above diagonal parts of the matrices show the measured correlation matrices for 
the dark matter (panel (a)), and the haloes in samples 1, 2 and 3 (panels (b), (c) and (d) respectively).
The lower triangular parts show the theoretical predictions. In general, Eq.~(\ref{eq:cov_xi2}) 
accurately describes the elements close to the diagonal, particularly for the dark matter. 
Further away from the diagonal, $Cor(i,j)$ is noisy and the agreement is not so clear.

\citet{smith08} performed a similar comparison and reach the same overall
conclusion as we do. These authors applied Eq.~(\ref{eq:cov_xi}) to describe 
the covariance matrix of the bin averaged correlation function in the limit 
$\Delta r/r \ll 1$. In order to include the effect of the bin width, they 
re-wrote the contribution to the covariance coming from the term involving  
$1/{\bar n}^2$ as 
\begin{equation}
 \frac{2}{V\,\bar{n}^2}\int \frac{ {\rm d}^3k}{(2\pi)^3}j_0(kr_i)j_0(kr_j)=\delta^K_{i,j}\frac{2}{V\,\bar{n}^2\,V_i},
\label{eq:aprox_smith}
\end{equation}
which corresponds to the exact result from Eq.~(\ref{eq:cov_xi2}) in the Poisson clustering 
limmit ($P(k)\ll1/\bar{n}$).
This correction includes the effect of binning only in the last term of 
Eq.~(\ref{eq:cov_xi}) and ignores the effect of the shell-average in the 
first two terms. Fig.~\ref{fig:variance} shows a comparison of 
the variance measured from our ensemble of simulations (solid line histogram) 
for the halo correlation functions in halo samples 1, 2 and 3 with the 
predictions from Eqs.~(\ref{eq:cov_xi2}) and~(\ref{eq:aprox_smith}). It can 
be clearly seen that even for the value of $\Delta r/r =0.03$, 
Eq.~(\ref{eq:aprox_smith}) slightly overpredicts the true variance, 
while the full expression from Eq.~(\ref{eq:cov_xi2}) gives a more accurate 
description.

Eq.~(\ref{eq:cov_xi2}) is very useful for predicting the error and full covariance matrix 
of two-point correlation functions estimated from galaxy samples. This is a valuable tool 
to use in forecasts of the likely constraints attainable on cosmological parameters from 
present and future galaxy surveys.

\begin{figure}
\includegraphics[width=0.47\textwidth]{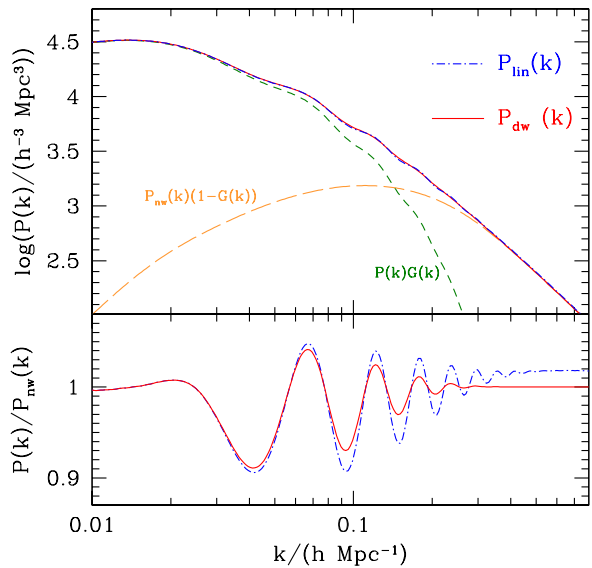}
\caption{
Upper panel: the linear perturbation theory power spectrum, 
$P_{\rm lin}(k)$ (dot-dashed line) compared with the ``de-wiggled'' 
power spectrum, $P_{\rm dw}(k)$ (solid line), as defined by 
Eq.~(\ref{eq:pdw}). The individual terms on the right hand side of 
Eq.~(\ref{eq:pdw}) which make up the dewiggled power spectrum 
are also shown by short and long dashed lines as labelled. 
Lower panel: the ratios of the linear (dot-dashed line) and 
dewiggled power spectra (solid line) to a smooth, cold dark matter 
only power spectrum$P_{\rm nw}(k)$. 
}
\label{fig:dewiggled}
\end{figure}

\section{Modelling the full shape of $\xi(r)$}
\label{sec:modelling_full_shape}

\subsection{Non-linear evolution}
\label{sec:non-linear}

The nonlinear evolution of density perturbations changes the shape of the power spectrum 
and the correlation function due to the coupling between different Fourier modes.
Numerical simulations have been used to model this effect \citep{eftathiou1988,hamilton1991,
pd1994,smith2003}. Based on a combination of the \citet{hamilton1991} scaling relations 
and the halo model, \citet{smith2003} proposed a fitting function to describe the effects 
of non-linear evolution on the shape of the matter power spectrum. This formula is able 
to reproduce the non-linear power spectra for pure CDM models to an accuracy of around 7\% 
and is the basis of the commonly used {\tt halofit} code. 

\citet{huff07} and \citet{crocce08} showed that 
the behaviour of the non-linear power spectrum predicted by {\tt halofit} 
fails to describe an important distortion to the baryonic acoustic oscillations 
seen in numerical simulations. Besides the change in the overall shape of $P(k)$, 
non-linear evolution washes out the acoustic oscillations by erasing the higher 
harmonic peaks \citep{meiksin1999,eisenstein05,springel2005,eisenstein06,angulo05,angulo08}.
This effect can be modelled by computing a damped or `dewiggled' power spectrum 
\citep{tegmark06,eisenstein06}:
\begin{equation}
P_{\rm dw}(k)=P_{\rm lin}(k)G(k)+P_{\rm nw}(k)(1-G(k)),
\label{eq:pdw}
\end{equation}
where $P_{\rm lin}(k)$ is the linear theory power spectrum and $P_{\rm nw}(k)$ 
is a smooth, linear theory, cold dark matter only power spectrum, with the 
same shape as $P_{\rm lin}(k)$ but without any baryonic oscillations 
(which can be computed using the fitting formula of \citet[]{EH99}). 
The function $G(k)\equiv \exp\left[ -(k/\sqrt{2}k_{\star})^2\right] $ regulates 
the transition from large scales ($k\ll k_{\star}$), where $P_{\rm dw}(k)$ 
follows linear theory to small scales ($k\gg k_{\star}$) where the acoustic 
oscillations are completely damped. The upper panel of Fig.~\ref{fig:dewiggled} 
compares $P_{\rm lin}(k)$ (dashed line) and $P_{\rm dw}(k)$ (solid line) together 
with the different terms in Eq.~(\ref{eq:pdw}). The damping of the BAO in the 
final power spectrum can be more clearly seen in the lower panel of 
Fig.~\ref{fig:dewiggled} which shows the ratios of the linear and dewiggled 
power spectra to $P_{\rm nw}(k)$.

The degree of damping of the BAO is extremely sensitive to the value 
adopted for $k_{\star}$. We treat $k_{\star}$ as a free parameter in the 
model for the form of the correlation function. Some authors have attempted 
to compute the value of $k_{\star}$. Based on the analysis of 
\citet{eisenstein06}, \citet{tegmark06} computed the value of the 
damping scale as a function of $\Omega_{\rm m}$ and the primordial 
amplitude of scalar fluctuations, $A_{\rm s}$. These authors expressed 
their results in terms of 
$\sigma_{\star}$ (where  $k_{\star}=1/\sigma_{\star}$):
\begin{equation}
\sigma_{\star}=s_0D(1+f)^{1/3}\sqrt{\frac{A_s}{0.6841}}.
\label{eq:kstar}
\end{equation}
Here, $s_0 = 12.4\,h^{-1}\,{\rm Mpc}$ is a reference scale, 
$D$ is the growth factor (which is where $\Omega_{\rm m}$ enters), 
$f={\rm d}\log D/{\rm d} \log a  $, and $A_s$ follows the normalisation 
convention of \citet{spergel07}.

\begin{figure}
\includegraphics[width=0.47\textwidth]{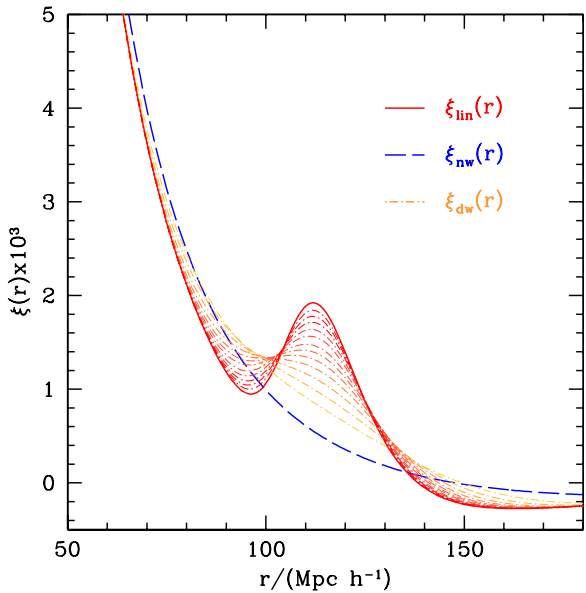}
\caption{
An illustration of how the appearance of the correlation function changes 
after applying differing degrees of damping to the acoustic oscillations. 
The undamped, linear theory correlation function is shown by the solid 
line. The damped or ``dewiggled'' correlation functions are shown by 
the dot-dashed lines, for different values of $k_{\star}$. For small damping 
(large $k_{\star}$), $\xi_{\rm dw}(r)$ deviates by a small amount from 
$\xi_{\rm lin}(r)$. As the damping of the acoustic oscillations becomes stronger 
(smaller $k_{\star}$), the peak is gradually erased and approaches the 
zero-baryon or ``no-wiggle'' correlation function, $\xi_{\rm nw}(r)$ (dashed line) 
as $k_{\star}\rightarrow0$.
}
\label{fig:kest}
\end{figure}

\begin{figure}
\includegraphics[width=0.47\textwidth]{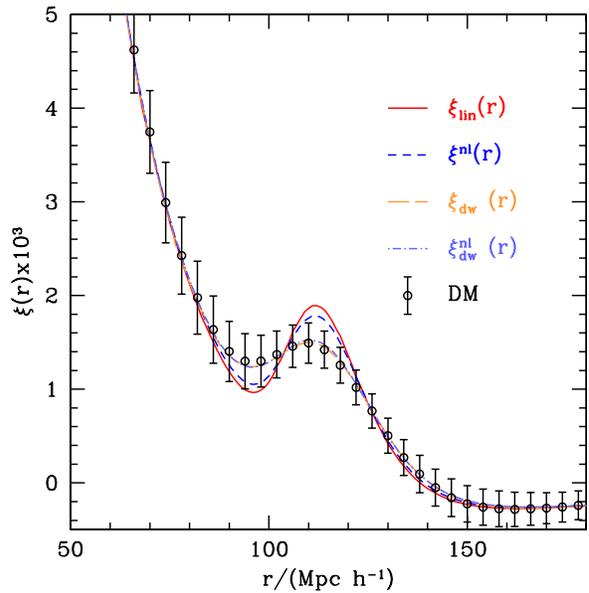}
\caption{
Comparison of the $z=0$ real-space dark matter two-point correlation function 
averaged over the ensemble of simulations (open points) with: (i) the linear 
theory correlation function $\xi_{\rm lin}(r)$ (solid line), (ii) an estimate of 
the non-linear correlation function $\xi^{\rm nl}(r)$ computed using 
{\tt halofit} without damping of the acoustic oscillations (dashed line), 
(iii) the dewiggled linear theory correlation function $\xi_{\rm dw}(r)$ 
defined by Eq.~(\ref{eq:xidw}) (dot-dashed line) and (iv) a dewiggled 
correlation function after being non-linearised using {\tt halofit} 
$\xi^{\rm nl}_{\rm dw}(r)$ (long-dashed line). 
The errorbars indicate the variance between the different realizations in 
the ensemble of simulations.
}
\label{fig:xi_nl}
\end{figure}

Eq.~(\ref{eq:pdw}) gives a purely phenomenological description of the 
damping of the acoustic oscillations found in numerical simulations. 
Using renormalized perturbation theory (RPT), \citet{crocce06,crocce08} 
give a theoretical justification of the ability of Eq.~(\ref{eq:pdw}) 
to describe non-linear evolution. According to RPT, the first term on 
the right hand side of Eq.~(\ref{eq:pdw}) describes the growth of a single 
mode, quantified by the propagator function $G(k)$. In the high-$k$ limit the propagator
is given by the Gaussian form with $k_{\star}$ given by \citep{crocce06,matsubara08}
\begin{equation}
k_{\star}= \left[ \frac{1}{3\pi^2} \int {\rm dk} P_{\rm lin}(k) \right] ^{-1/2}.
\label{eq:kstar_rpt}
\end{equation}
In the next section we shall compare the value of $k_{\star}$ obtained from 
Eqs.~(\ref{eq:kstar}) and~(\ref{eq:kstar_rpt}) with the best 
fitting value obtained by requiring our model for the correlation function 
to reproduce measurements made from the N-body simulations.

The second term on the right hand side of Eq.~(\ref{eq:pdw}) can be interpreted as
the power generated by the coupling of Fourier modes on 
small scales, $P_{\rm mc}(k)$. The term $P_{\rm mc}(k)$ is negligible on 
large scales (small $k$), but dominates the total power on small scales 
(high $k$). For the scales relevant to the analysis of BAO, which are of 
the order of $k_{\star}$, $P_{\rm mc}$ has a similar amplitude to 
$P_{\rm nw}(k)(1-G(k))$. 

Note, however, that according to RPT the propagator $G(k)$ only behaves 
as a Gaussian in the high-$k$ limit. Besides $P_{\rm nw}(k)(1-G(k))$ is a 
smooth function while, as shown by \citet{crocce08}, the term $P_{\rm mc}$ 
does show acoustic oscillations, although of a much smaller amplitude than $P(k)$. 
This differences constitute the limitation of a modelling based on Eq.~(\ref{eq:pdw}). 
We will return to this point in \S~\ref{sec:imp}.

\begin{figure}
\includegraphics[width=0.44\textwidth]{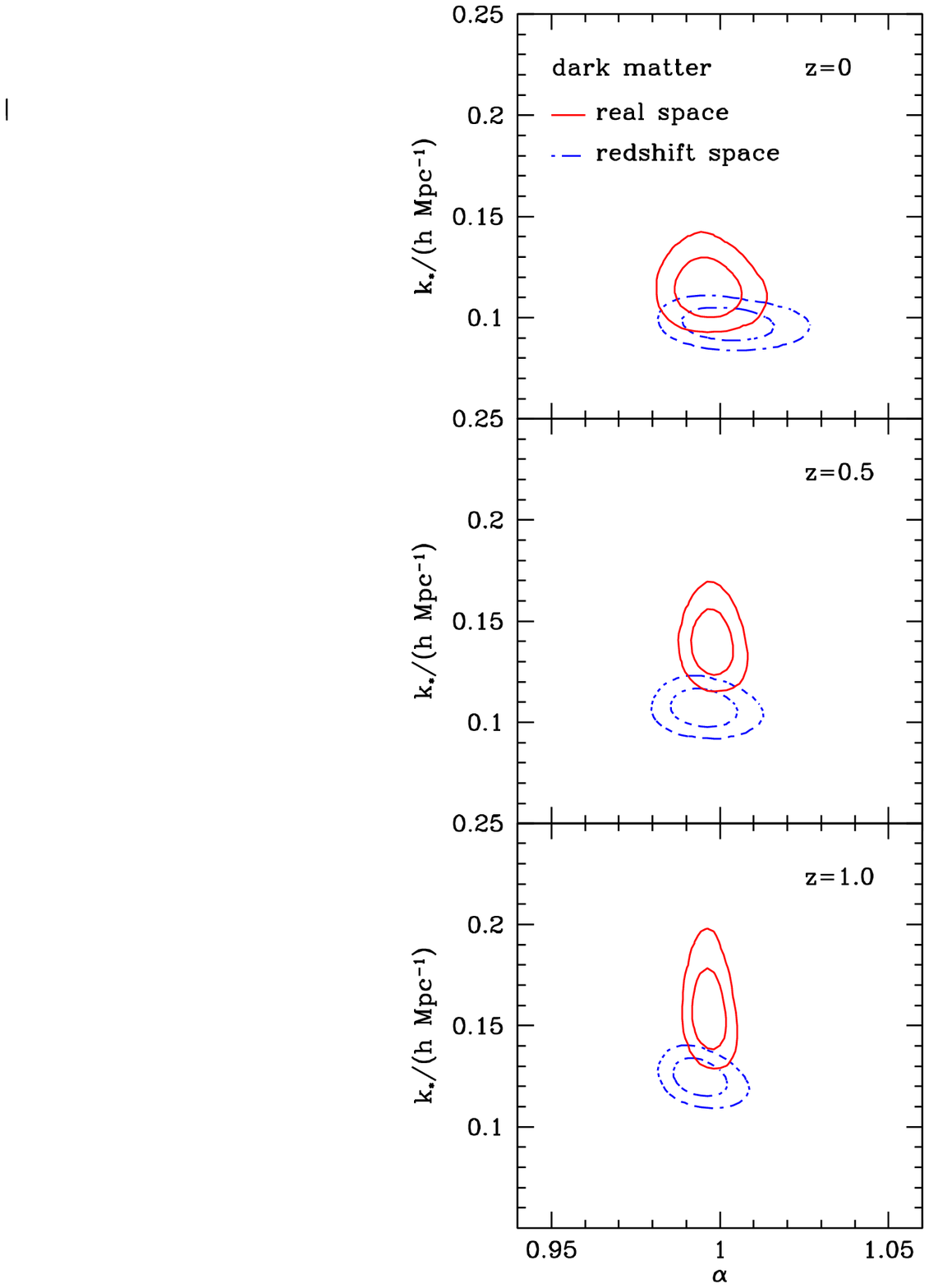}
\caption{
Constraints on $\alpha$ and $k_{\star}$ obtained using the mean 
real-space (solid lines) and redshift-space (dashed lines) 
dark matter correlation functions from the ensemble of simulations 
at redshift $z=0$, 0.5 and 1 (panels (a), (b) and (c) respectively).
}
\label{fig:cont_dm}
\end{figure}

Taking the Fourier transform of Eq.~(\ref{eq:pdw}), 
the two-point correlation function will be given by
\begin{equation}
\xi_{\rm dw}(r)=\xi_{\rm lin}(r)\otimes \tilde{G}(r) + 
\xi_{\rm nw}(r)\otimes(1-\tilde{G}(r)),
\label{eq:xidw}
\end{equation}
where the symbol $\otimes$ denotes a convolution and $\tilde{G}(r)$ 
is the Fourier transform of $G(k)$. The information about the acoustic 
oscillations is contained in the first term, which represents the 
convolution of the linear theory correlation function with a Gaussian kernel. 
This convolution implies that in the correlation function, the damping 
of the higher harmonic oscillations causes the acoustic peak to broaden 
and shift to smaller scales \citep{smith08,crocce08}. Fig.~\ref{fig:kest} shows the dewiggled 
correlation functions, $\xi_{\rm dw}(r)$, obtained for different 
values of $k_{\star}$ (dot-dashed lines). For weak damping (large $k_{\star}$), 
only the highest harmonic oscillations are damped and $\xi_{\rm dw}(r)$ 
deviates only slightly from $\xi_{\rm lin}(r)$ (solid line). As the damping 
becomes stronger (smaller $k_{\star}$), the impact on the shape of the peak 
increases, with $\xi_{\rm dw}(r)$ approaching $\xi_{\rm nw}(r)$ (dashed line) 
as $k_{\star}\rightarrow0$.

A full description of the non-linear two-point correlation function must 
include both the overall change in shape of the correlation function and 
the damping of the acoustic oscillations. In order to assess which of these 
effects is the more important for the analysis of the acoustic peak in 
$\xi(r)$, Fig.~\ref{fig:xi_nl} compares the mean $z=0$ real-space correlation 
function of the dark matter measured from the ensemble of simulations (open points) 
with the following models for the correlation function: 
(i) the linear theory correlation function $\xi_{\rm lin}(r)$ (solid line), 
(ii) a non-linear correlation function $\xi^{\rm nl}(r)$ computed using 
{\tt halofit}, without any damping of the acoustic oscillations (short-dashed line), 
(iii) the dewiggled linear theory correlation function $\xi_{\rm dw}(r)$, computed 
as described by Eq.~(\ref{eq:xidw}) (long-dashed line) 
and (iv) a dewiggled correlation function non-linearised using {\tt halofit} 
$\xi^{\rm nl}_{\rm dw}(r)$ (dot-dashed line). 
The errorbars indicate the variance between the correlation functions measured 
from the different realizations in the simulation ensemble. 

It is clear from Fig.~\ref{fig:xi_nl} that the acoustic peak in the two-point 
correlation function at redshift $z=0$ shows strong deviations from the 
predictions of linear theory. The {\tt halofit} fitting formula fails to describe 
these deviations since its application only produces a small decrease in the 
amplitude of the peak without shifting its position. On the contrary, the linear 
theory dewiggled correlation function from Eq.~(\ref{eq:xidw}) gives a very good 
description of the results of our numerical simulations, showing that the damping 
of the oscillations is the most important effect to include in the modelling of 
the real space correlation function on large scales. The incorporation of the full 
change in shape of $P(k)$ due to non-linear evolution produces very little 
difference in the shape of the acoustic peak in the correlation function. However, 
as we will see later (see section~\ref{sec:imp}), this effect might be 
important on intermediate scales ($r \simeq 70\,h^{-1}\,{\rm Mpc}$).

\subsection{The model in practice}
\label{sec:fitting}

\begin{figure}
\includegraphics[width=0.47\textwidth]{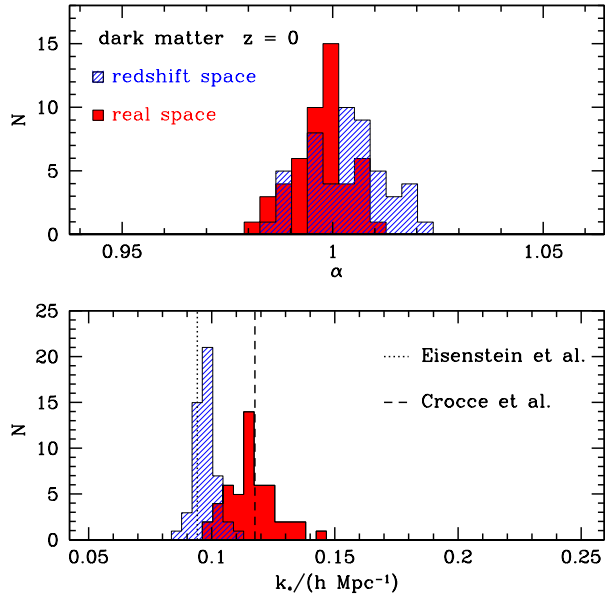}
\caption{
The distribution of the values of $\alpha$ (upper panel) and $k_{\star}$ 
(lower panel) recovered by fitting the model of Eq.~(\ref{eq:xidw}) to the $z=0$
correlation functions measured from the different realizations in real (solid histogram) 
and redshift-space (shaded histogram). The dotted and dashed lines in the lower panel 
show the value of $k_{\star}$ predicted by Eq.~(\ref{eq:kstar}) and (\ref{eq:kstar_rpt}) respectively.
}
\label{fig:histo_a_k}
\end{figure}

We now test whether or not the model for the correlation function 
described in \S~\ref{sec:non-linear} returns unbiased constraints 
on cosmological parameters. In particular, we are interested in the 
constraints on the dark energy equation of state parameter $w_{\rm DE}$. 
We consider a very simple case in which we assume that the values of 
all cosmological parameters are known, apart from $w_{\rm DE}$, and 
analyse the constraints on this parameter. 

\begin{figure*}
\centering
\centerline{\includegraphics[width=0.97\textwidth]{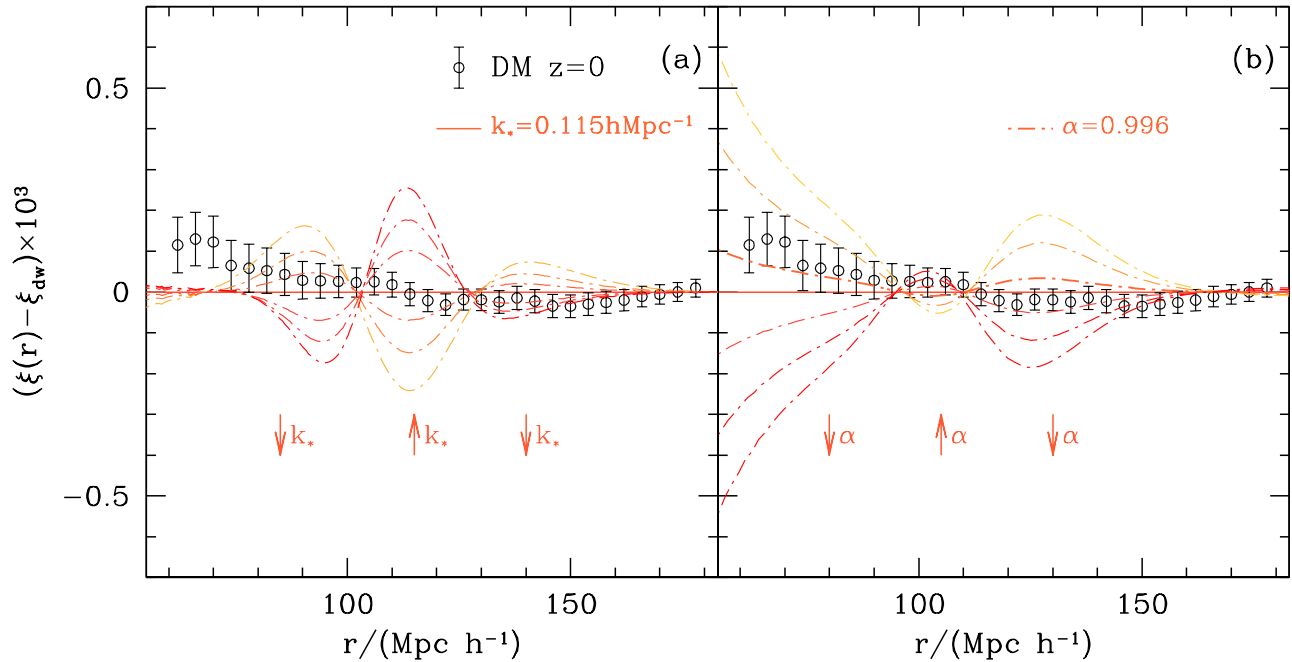}}
\caption{
Panel (a): dewiggled correlation functions corresponding to $k_{\star}=0.09$, 0.1, 0.11, 0.15. 0.17 and 0.21, for a fixed
stretch parameter, $\alpha=1$ (dot-dashed lines). To increase the dynamic range, we show the residuals
$\xi(r)-\xi_{\rm dw}(r)$, where $\xi_{\rm dw}(r)$ is the dewiggled correlation function for the best fit
value of $k_{\star}=0.115 \,h\,{\rm Mpc^{-1}}$.
The most important deviations between $\xi_{\rm dw}$ and the measured correlation function is found at
scales smaller than the position of the peak, with $r \lesssim 90\,h^{-1}{\rm Mpc}$.
Panel (b): Difference $\xi(r)-\xi_{\rm dw}(r)$ for $\alpha=0.978$, 0.986, 0.996, 1.006, 1.014 and 1.022, fixing $k_{\star}=0.115 \,h\,{\rm Mpc^{-1}}$ (dot-dashed lines). A value of $\alpha$ slightly less than 1 re-scales the position of the peak in the model correlation function and improves the agreement with the measurements in small scales, causing the slight bias of the model thowards $\alpha<1$.
The arrows indicate the effect of increasing $k_{\star}$ and $\alpha$ on the model correlation functions.
The errorbars show the variance in the mean correlation function, that is $\sigma_{\bar{\xi}}=\sigma_{\xi}/\sqrt{50}$.
}
\label{fig:diff}
\end{figure*}

The change in the power spectrum due to variations in $w_{\rm DE}$ 
in this simple case is described in detail by \citet{angulo08}. 
In order to measure the power spectrum of galaxy clustering, it is 
necessary to convert the angular positions and redshifts of the galaxies 
into comoving spatial separations. This requires a choice to be made 
for the values of the cosmological parameters, including $w_{\rm DE}$. 
The effect of a change in the value of $w_{\rm DE}$ from its true value 
to $w_{\rm DE}+\delta w_{\rm DE}$ is to modify the separations between 
pairs of galaxies, which leads to a change in the appearance of the 
power spectrum. For small perturbations away from the true equation of 
state, the alteration in the measured power spectrum can be represented by a
rescaling of the wavenumber from $k_{\rm true}$ to $k_{\rm app}$. This 
change can be described by a `stretch' factor $\alpha$, defined by \citep{huff07,angulo08}
\begin{equation}
 \alpha=\frac{k_{\rm app}}{k_{\rm true}}.
\end{equation}
In the two-point correlation function, there will be an equivalent shift from 
scale $r_{\rm true}$ to $r_{\rm app}=r_{\rm true}/\alpha$.
We note that the variant models which arise from perturbing the equation of 
state in this way do not match observations such as the location of the Doppler 
peaks in the cosmic microwave background \citep[see][]{angulo08}. Nevertheless, 
our primary goal here is to compare the constraints on the stretch parameter obtained 
from the correlation function with those which result from the power spectrum, under 
the same idealized conditions.

We analyse the constraints on the stretch parameter using the two-point 
correlation functions in real-space measured from the different {\tt L-BASIIC}~{\tt II} 
realizations, as described in Section~\ref{sec:estimator}, as well as from the 
mean correlation function from the ensemble. In doing so, we use only information 
from the shape of the correlation function, and not its overall amplitude, which for real 
observational data can be affected by bias and redshift-space effects (see 
Sections~\ref{sec:redshift} and~\ref{sec:bias}). We also consider $k_{\star}$ as a free 
parameter, instead of fixing its value according to Eq.~(\ref{eq:kstar}). To explore 
this simple parameter space we constructed Monte Carlo Markov Chains (MCMC). The 
constraints obtained in this way for redshifts $z=0$, 0.5 and 1 are summarised in 
Table~\ref{tab:params}. The range of pair separations used in the fit is 
$60 < (r/ h^{-1}{\rm Mpc}) < 180$; the results are not sensitive in detail to the 
choice of limits in $r$.
Unless otherwise stated, the quoted allowed ranges for the constrained parameters 
correspond to the $68\%$ confidence level according to the variance from our ensemble
of simulations.

The solid lines in Fig.~\ref{fig:cont_dm} show the two-dimensional constraints 
on $\alpha$ and $k_{\star}$ obtained using the mean dark matter correlation 
function in real-space measured from the ensemble of simulations. The constraints 
on $k_{\star}$ can be compared with the predictions of Eq.~(\ref{eq:kstar})
which for this cosmology gives $k_{\star}=0.095\,h\,{\rm Mpc^{-1}}$ at $z=0$. 
Here we find that at $z=0$, the shape of the mean correlation function can be 
more accurately described by $k_{\star}=0.115 \pm 0.009\,h\,{\rm Mpc^{-1}}$, 
which is $2-\sigma$ away from the prediction of Eq.~(\ref{eq:kstar}). 
On the other hand, this result is in excellent agreement with the prediction from Eq.~(\ref{eq:kstar_rpt})
which gives $k_{\star}=0.117\,h\,{\rm Mpc^{-1}}$ at $z=0$.
The lower panel of Fig.~\ref{fig:histo_a_k} shows the histogram of the values of 
$k_{\star}$ obtained for the $z=0$ correlation functions of each {\tt L-BASIIC}~{\tt II} 
realization, which is completely consistent with the constraints obtained from 
the mean correlation function of the ensemble.
The dotted line shows the value of $k_{\star}$ computed using Eq.~(\ref{eq:kstar}) which, 
as we mentioned before, predicts a stronger damping of the acoustic oscillations than 
we find in the numerical simulations for this cosmology. This implies that if one applies   
Eq.~(\ref{eq:kstar}), this may result in systematic errors in the values of the derived 
cosmological parameters. 
The dashed line shows the prediction from Eq.~(\ref{eq:kstar_rpt}) which is remarkably
consistent with the mean value of $k_{\star}$ recovered from our ensemble of simulations.

At higher redshift, the acoustic oscillations are less damped and higher values of 
$k_{\star}$ (corresponding to weaker damping) are favoured, with 
$k_{\star}=0.140 \pm 0.010\,h\,{\rm Mpc^{-1}}$ returned at $z=0.5$ and 
$k_{\star}=0.159 \pm 0.013\,h\,{\rm Mpc^{-1}}$ at $z=1$. The constraints on $k_{\star}$ 
also broaden since, as can be seen in Fig.~\ref{fig:xi_diff}, for higher values of 
$k_{\star}$ the shape of the acoustic peak becomes less sensitive to variations in 
this parameter.

At redshift $z=0$, the mean correlation function gives tight constraints on the stretch 
factor, with $\alpha=0.996\pm 0.006$, showing that it is better described with a value 
of $\alpha$ slightly less than 1. The upper panel of Fig.~\ref{fig:histo_a_k} shows the 
histogram of the values of $\alpha$ recovered from the different realizations, which is 
centred on the value $\alpha=1$, but which also shows a small tendency towards $\alpha<1$ 
which can be detected by the slight skewness of the distribution. At higher redshifts the 
position of the acoustic peak can be more precisely determined since it is less affected 
by the damping of the higher harmonic oscillations. This results in tighter constraints 
on $\alpha$, with $\alpha=0.998\pm 0.004$ at $z=0.5$ and $\alpha=0.997\pm 0.003$ at $z=1$. 

The origin of this slight bias towards $\alpha<1$ can be understood by
analysing the effect of variations in $k_{\star}$ and $\alpha$ on the model correlation function.
Fig.~\ref{fig:diff}(a) shows the correlation functions corresponding to different 
values of $k_{\star}$ for a fixed stretch parameter, $\alpha=1$ (dot-dashed lines).
To increase the dynamic range, we plot the residuals $\xi(r)-\xi_{\rm dw}(r)$, where $\xi_{\rm dw}(r)$
is the dewiggled correlation function for the best fit value of $k_{\star}=0.115 \,h\,{\rm Mpc^{-1}}$.
For this reference value, $\xi_{\rm dw}$ accurately describes the shape of the acoustic peak in the mean 
correlation function measured from the ensemble of simulations. There are, however, 
differences between the model of Eq.~(\ref{eq:xidw}) and the measured correlation function 
both on scales smaller and larger than the acoustic peak which present a systematic structure.

At scales around $r \simeq 150\,h^{-1}{\rm Mpc}$, Eq.~(\ref{eq:xidw}) slightly 
overestimates the true value of the correlation function, which follows more closely 
the result from linear perturbation theory. On these scales, the value of $\xi_{\rm dw}$ 
is almost completely determined by the first term of Eq.~(\ref{eq:xidw}), with the 
second term giving only a slight correction. This difference may indicate that the 
function $G(k)$ that controls the damping of the oscillations is not exactly Gaussian, 
thereby changing the shape of the first term of Eq.~(\ref{eq:xidw}), or that the second 
term in this equation is not a good approximation to $\xi_{\rm mc}(r)$, and underestimates  
the full contribution of this term.

The most important deviations between the model and the measured correlation function 
are found on scales smaller than the position of the peak, with $r \lesssim 90\,h^{-1}
{\rm Mpc}$. On these scales the model underestimates the amplitude of the measured 
correlation function. 

Fig.~\ref{fig:diff}(b) shows the correlation functions obtained for different values 
of $\alpha$, fixing $k_{\star}=0.115 \,h\,{\rm Mpc^{-1}}$ (dot-dashed lines). As stated 
before, for $\alpha=1$ the model fails to reproduce the measured  
correlation function. Using a value of $\alpha$ slightly less than 1 re-scales the position 
of the peak and increase the amplitude at small scales in the model correlation function in 
a way that gives better agreement with the measurements from the simulations, causing the
slight bias of the model towards $\alpha<1$.

\citet{crocce08} showed that $\xi_{\rm mc}(r)$ introduces an additional shift in the 
acoustic peak which is not taken into account by the second term in the right hand side of Eq.(\ref{eq:xidw}).
At $z=0$ this shift can lead to a bias towards $\alpha > 1$ of the order of $0.8\%$ if 
this parameter is fitted using only the position of the peak (see their Figure 7). 
This apparent contradiction with our result of a small bias towards $\alpha < 1$ 
is due to the fact that our constraints are based on the full shape of the 
correlation function on large scales, and not only on the position of the peak. 
In this case, as we shown above, the biggest difference between the models and 
the measured correlation functions is not given at the scales of the peak, but 
at smaller scales, where non-linear evolution increases the amplitude of $\xi(r)$ 
with respect to the prediction of Eq.(\ref{eq:xidw}). This difference is responsible
for the shift in the preferred value of $\alpha$.

\subsection{Improving the model}
\label{sec:imp}

Although the model described by Eqs.~(\ref{eq:pdw}) and (\ref{eq:xidw}) gives a 
good description of the distortions in the shape of the peak in the correlation function 
due to the damping of the acoustic oscillations, it is slightly biased towards $\alpha<1$.
This bias is smaller than the sample variance predicted in the correlation function for
the volume of the {\tt L-BASIIC} simulation or expected in ongoing galaxy surveys.
However, the much smaller sample variance anticipated in future galaxy surveys
such as Euclid \citep{SPACE} or ADEPT,
will mean that this level of inaccuracy in the modelling of the correlation 
function may lead to the introduction of systematic errors in the recovered dark energy 
equation of state parameter $w_{\rm DE}$. In this section we explore different ways 
to correct for this small bias.

\begin{figure*}
\includegraphics[width=0.97\textwidth]{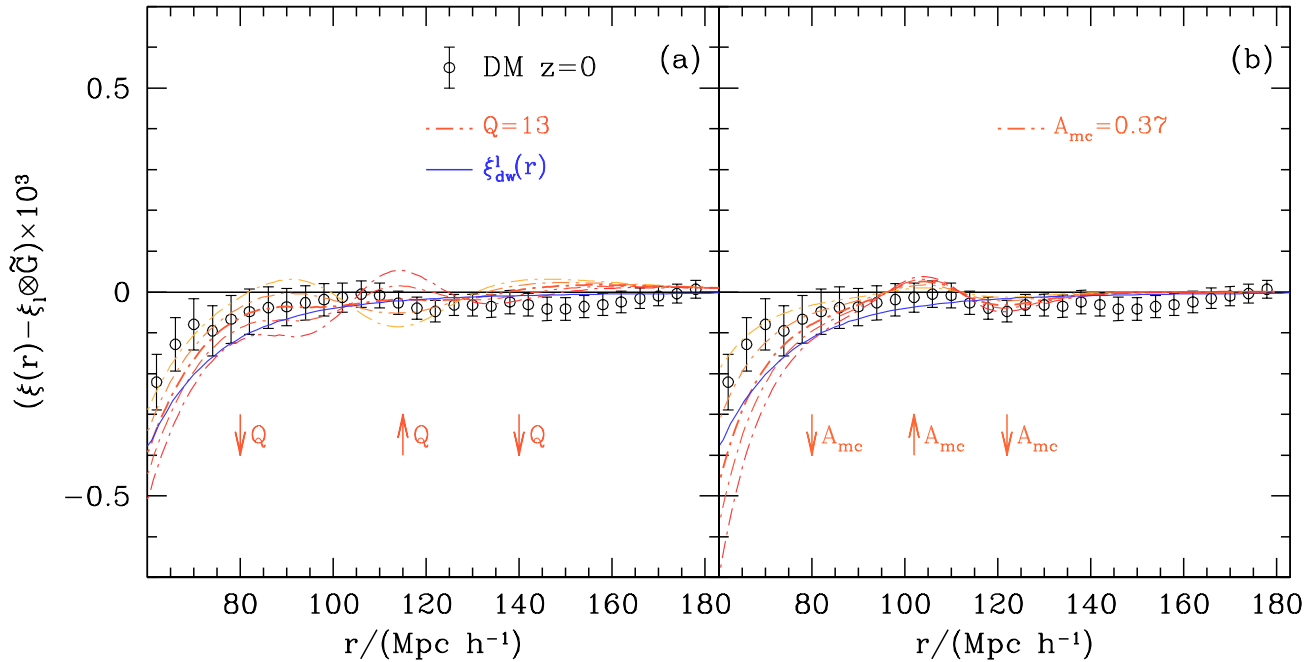}
\caption{
Panel (a): The quantity $(\xi^{\rm nl}_{\rm dw}-\xi_{\rm lin}\otimes \tilde{G})$ for $Q=1$, 7, 13, 19 and 26, fixing $\alpha=1$ and 
$k_{\star}=0.115 \,h\,{\rm Mpc^{-1}}$ (dot-dashed lines). For $Q=13$ (thick dot-dashed line),
 the non-linear correction factor from Eq.~(\ref{eq:pdw_nl}) changes the shape of the model 
correlation function, improving the agreement with the measurements from the simulations 
on scales $r \lesssim 90\,h^{-1}{\rm Mpc}$. 
Panel (b): The quantity $(\xi^{\rm nl}-\xi_{\rm lin}\otimes \tilde{G})$ computed using Eq.~(\ref{eq:xinl_rpt}) with $A_{\rm mc}=0.15$, 0.25, 0.37, 0.45 and 0.55, fixing $\alpha=1$ and $k_{\star}=0.115 \,h\,{\rm Mpc^{-1}}$ (dot-dashed lines). 
This modelling gives a better description of the residuals of the measured correlation functions with respect to
 $(\xi_{\rm lin}\otimes \tilde{G})$ close to the acoustic peak.
The solid lines on both panels show the difference $(\xi^{\rm nl}_{\rm dw}-\xi_{\rm lin}\otimes \tilde{G})$, that is,
the second term on the right hand side of Eq.~(\ref{eq:xidw}). The arrows indicate the effect of increasing 
$Q$ and $A_{\rm mc}$ on the model correlation functions. The errorbars show the variance in the mean correlation
function, that is $\sigma_{\bar{\xi}}=\sigma_{\xi}/\sqrt{50}$.
}
\label{fig:q_diff}
\end{figure*}

The model of Eq.~(\ref{eq:pdw}) can be extended to account for the change in the overall 
shape of the power spectrum due to non-linear evolution by
\begin{equation}
P_{\rm dw}^{\rm nl}(k)=\left( \frac{1+Q\,k^2}{1+Ak+Bk^2}\right)P_{\rm dw}(k) 
= f(k) P_{\rm dw}(k). 
\label{eq:pdw_nl}
\end{equation}
The factor $f(k)$ could also be used to model a scale dependent bias factor. 
This model for non-linear evolution is based on the $Q$-model of \citet{cole05}, 
modified by the addition of a new parameter, $B$, with the aim of achieving a better 
description of the behaviour of the non-linear power spectrum at high $k$. 
The value of the new parameter $B$ depends on the choice of $Q$. In practice, 
we fixed the value of $B$ using the relation $B=Q/10$, which gives the approximate 
behaviour of the non-linear power spectrum at large $k$, although the correlation 
function is almost insensitive to this choice.
This scheme is an alternative means to describe the 
non-linear growth to using {\tt halofit}. 
This expression can be Fourier transformed to 
obtain a non-linear correlation function $\xi^{\rm nl}_{\rm dw}(r)$, whose shape is only 
weakly dependent on the precise value of $B$.

Fig.~\ref{fig:q_diff}(a) shows the quantity $\xi^{\rm nl}_{\rm dw}-\xi_{\rm l}\otimes \tilde{G}$ 
for different values of the parameter $Q$ fixing $\alpha=1$ and 
$k_{\star}=0.115 \,h\,{\rm Mpc^{-1}}$ (dot-dashed lines). For $Q=13$ (thick dot-dashed line),
 the non-linear correction factor from Eq.~(\ref{eq:pdw_nl}) changes the shape of the model 
correlation function, improving the agreement with the measurements from the simulations 
on scales $r \lesssim 90\,h^{-1}{\rm Mpc}$. On large scales the measurements follow more 
closely the predictions from linear perturbation theory and the models overpredict
the amplitude of the correlation function. 

Table~\ref{tab:params} also shows the constraints on $\alpha$ and $k_{\star}$ obtained 
by including the scale dependent correction factor defined by Eq.~(\ref{eq:pdw_nl}), on  
allowing $Q$ to vary as a free parameter. As the model correlation function is fairly 
insensitive to the value of $Q$ the constraints on this parameter are weak. However,  
there is a tendency towards lower values of $Q$ with increasing redshift.
As can be seen in Fig.~\ref{fig:q_diff}, varying $Q$ changes the amplitude of the 
acoustic peak. This is why when this parameter is included in the analysis, the best 
fits shift towards lower values of $k_{\star}$. This reflects the fact that when the non-linear correction 
of Eq.~(\ref{eq:pdw}) is not included, the value of $k_{\star}$ recovered from the 
correlation function contains some information on the distortion to the overall shape 
of $P(k)$ and not just the way in which the oscillations are damped.

The constraints on $\alpha$ are essentially unchanged upon the incorporation of the 
non-linear shape correction. The best fit results, as before, show a small bias 
towards $\alpha<1$. Hence, although the model presented in Eq.~(\ref{eq:pdw_nl}) 
gives the freedom to correct for differing degrees of non-linear evolution or even 
a scale dependence of bias, it is still unable to describe the form of the measured 
correlation functions to correct for this small systematic shift in $\alpha$.

The solid curve in Fig.~\ref{fig:q_diff} shows the quantity  
$\xi^{\rm nl}_{\rm dw}-\xi_{\rm l}\otimes \tilde{G}$, that is, 
the second term on the right-hand side of Eq.~(\ref{eq:xidw}). 
As pointed out in \S~\ref{sec:modelling_full_shape}, this term 
is a smooth function with no information about the acoustic peak. 
On the other hand, \citet{crocce08} showed that the term 
$\xi_{\rm mc}(r)$ does contains information about the acoustic scale. 
This fact imposes a limitation on the ability of a model based on 
Eq.~(\ref{eq:xidw}) to reproduce the full shape of the non-linear 
correlation function.

Using standard perturbation theory, \citet{crocce08} showed that 
on the scale of the acoustic peak the main contribution to 
$\xi_{\rm mc}(r)$ will be of the form
\begin{equation}
 \xi_{\rm mc}(r) \propto \xi'_{\rm lin}\,\xi^{(1)}_{\rm lin}(r),
\label{eq:ximc}
\end{equation}
where $\xi'_{\rm lin}$ is the derivative of the linear theory correlation function and
\begin{equation}
 \xi_{\rm lin}^{(1)}(r) \equiv \hat{r} \cdot \nabla^{-1}\xi_{\rm lin}(r)=4\pi\,\int P_{\rm lin}(k)\,j_1(kr)k\,{\rm d}k. 
\label{eq:xi1}
\end{equation}
Based on this result, Crocce \& Scoccimarro proposed an approximate phenomenological 
model to describe the effect of the mode coupling on the correlation function near 
to the acoustic scale, in which $\xi_{\rm mc}(r)\approx A_{\rm mc}\,\xi'_{\rm lin}$ 
(since  $\xi^{(1)}_{\rm lin}(r)$ varies little on these scales). We tested this 
ansatz by comparing the results from our simulations against the non-linear 
correlation function given by
\begin{equation}
 \xi_{\rm nl}(r) =\xi_{\rm lin}(r)\otimes \tilde{G}(r) + A_{\rm mc}\,\xi'_{\rm lin}\xi^{(1)}_{\rm lin}(r),
\label{eq:xinl_rpt}
\end{equation}
where $A_{\rm mc}$ is a free parameter. Fig.~\ref{fig:q_diff}(b) shows the non-linear 
correlation functions obtained for different values of $A_{\rm mc}$ fixing $\alpha=1$ 
and $k_{\star}=0.115 \,h\,{\rm Mpc^{-1}}$ (dot-dashed lines). It can be seen that 
this model can describe the shape of the residuals of the measured correlation 
functions with respect to $\xi_{\rm lin}\otimes \tilde{G}$ close to the acoustic peak. 
At smaller scales, where the approximation is not so accurate, the model underestimates 
the correlation functions.

Table~2 also shows the results obtained by applying this modelling. The  
constraints obtained on $k_{\star}$ are slightly smaller than those obtained 
using Eq.~(\ref{eq:xidw}). The most important effect of the incorporation of 
the second term of Eq.(\ref{eq:xinl_rpt}) is that it helps to alleviate the problem of
the small bias towards $\alpha<1$ found in \S~\ref{sec:fitting} yielding 
$\alpha=1.003\pm0.008$ at $z=0$, with a slight increase in the allowed region. 
At higher redshifts the constraints are tighter, with $\alpha=1.002\pm0.005$
at $z=0.5$ and $\alpha=1.000\pm0.003$ at $z=1$, in better agreement with $\alpha=1$. 
The ability of this simple ansatz to improve the obtained constraints might indicate 
that the implementation of a full calculation of $\xi_{\rm mc}$ using RPT over the 
full range of scales included in the analysis could help to improve the constraints 
even further.

\begin{table*}
\begin{center}{
\small
\begin{tabular}{|c c c c | c c | c c}
\noalign{\smallskip}

\hline
\hline
\noalign{\smallskip}
\noalign{\smallskip}
    &     &  &     & \multicolumn{2}{c}{Real-space} & \multicolumn{2}{c}{Redshift-space}\\
\noalign{\smallskip}
 Sample id & $z$ & $b^2$ & $S$ & $k_{\star}/(h\,Mpc^{-1})$ & $\alpha$  & $k_{\star}/(h\,Mpc^{-1})$ & $\alpha$\\
\noalign{\smallskip}
\hline
\noalign{\smallskip}
    & 0.0 & 1.00 & 1.34 & $0.115 \pm 0.009$  & $0.996 \pm 0.006$ & $0.097 \pm 0.004$  & $1.002 \pm 0.009$ \\
DM  & 0.5 & 1.00 & 1.56 & $0.140 \pm 0.010$  & $0.998 \pm 0.004$ & $0.107 \pm 0.005$  & $0.995 \pm 0.006$ \\
    & 1.0 & 1.00 & 1.69 & $0.159 \pm 0.013$  & $0.997 \pm 0.003$ & $0.124 \pm 0.005$  & $0.994 \pm 0.005$\\
\hline
    & 0.0 & 2.81 & 1.19& $0.117 \pm 0.017$  & $0.997 \pm 0.012$ & $0.094 \pm 0.013$  & $0.992 \pm 0.012$\\
Halos 1  & 0.5 & 5.57 & 1.21 & $0.146 \pm 0.024$  & $0.995 \pm 0.011$ & $0.107 \pm 0.014$  & $1.001 \pm 0.013$ \\
    & 1.0 & 11.58 & 1.17 & $0.22 \pm 0.10$  & $0.997 \pm 0.014$ & $0.131 \pm 0.028$  & $0.997 \pm 0.017$\\
\hline
    & 0.0 & 3.40 & 1.17 &$0.170 \pm 0.065$  & $0.997 \pm 0.010$ & $0.099 \pm 0.024$  & $0.998 \pm 0.020$ \\
Halos 2  & 0.5 & 7.35 & 1.18 & $0.20 \pm 0.10$  & $1.006 \pm 0.015$ & $0.160 \pm 0.066$  & $1.002 \pm 0.020$ \\
    & 1.0 & 16.09 & 1.15 &$0.27 \pm 0.15$  & $1.003 \pm 0.022$ & $0.22 \pm 0.13$  & $0.998 \pm 0.019$ \\
\hline
    & 0.0 & 6.46  & 1.12 &$0.121 \pm 0.023$  & $0.994 \pm 0.012$ & $0.109 \pm 0.019$  & $0.990 \pm 0.015$\\
Halos 3  & 0.5 & 13.14& 1.13 & $0.203 \pm 0.076$  & $0.988 \pm 0.017$ & $0.187 \pm 0.070$  & $0.983 \pm 0.017$ \\
    & 1.0 & 28.13 & 1.11&$0.28 \pm 0.13$  & $1.000 \pm 0.025$ & $0.24 \pm 0.13$  & $1.000 \pm 0.021$ \\

\noalign{\smallskip}
\hline
\hline
\end{tabular}
\caption{
Results of applying the general fitting procedure described in Sec.~\ref{sec:non-linear} to
 the correlation function of the different samples analysed in real-space and redshift-space.
The first column gives the label of the sample, as defined in Section 2. The second column gives the redshift output. 
Column 3 gives the real-space bias factors computed as described in Sec.~\ref{sec:bias}. Column 4 gives the value of the redshift-space boost factor computed according to Eq.~(\ref{eq:xi_s}). Columns 5 and 6 give the constraints on $k_{\star}$ and $\alpha$ obtained by fitting $\xi_{\rm dw}$ as in Eq.~(\ref{eq:xidw}) in real-space while columns 7 and 8 show the results obtained using the redshift-space measurements. All quoted values correspond to the mean and 68\% c.l. of the respective parameter.
}
}\end{center}
\label{tab:params1}
\end{table*}

\begin{table*}
\begin{center}{
\small
\begin{tabular}{|c c | c c c | c c c |}
\noalign{\smallskip}
\noalign{\smallskip}
       \multicolumn{8}{c}{Real-space}  \\
\noalign{\smallskip}
\hline
\hline
\noalign{\smallskip}
\noalign{\smallskip}
    &      & \multicolumn{3}{c}{Fits with $\xi^{\rm nl}_{\rm dw}$} & \multicolumn{3}{c}{Fits with $\xi^{\rm nl}_{\rm mc}$}\\
\noalign{\smallskip}
 Sample id & $z$ &$k_{\star}/(h\,Mpc^{-1})$ & $\alpha$ &$Q$ & $k_{\star}/(h\,Mpc^{-1})$ & $\alpha$ & $A_{\rm mc}$   \\
\noalign{\smallskip}
\hline
\noalign{\smallskip}
    & 0.0 & $0.107 \pm 0.013$ & $0.996 \pm 0.006$ & $18 \pm 12$ & $0.1135\pm0.0089$ & $1.003\pm0.008$&$0.37\pm0.22$ \\
DM  & 0.5 & $0.126 \pm 0.018$ & $0.997 \pm 0.004$ & $19 \pm 12$ & $0.133\pm0.010$ & $1.002\pm0.005$&$0.39\pm0.21$\\
    & 1.0 & $0.155 \pm 0.030$ & $0.997 \pm 0.003$ & $13 \pm 10$ & $0.155\pm0.013$ & $1.000\pm0.003$&$0.27\pm0.18$\\
\hline
    & 0.0 & $0.115 \pm 0.021$ & $0.996 \pm 0.012$ & $16 \pm 11$ & $0.120\pm0.016$ & $1.002\pm0.013$ & $0.35\pm0.25$\\
Halos 1  & 0.5  & $0.151 \pm 0.032$ & $0.993 \pm 0.011$ & $14 \pm 11$ & $0.142\pm0.022$ &$0.999\pm0.012$ &$0.41\pm0.28$\\
    & 1.0   & $0.20 \pm 0.10$ & $0.995 \pm 0.013$ & $15 \pm 12$ &  $0.199\pm0.087$&$1.003\pm0.015$ &$0.52\pm0.36$\\
\hline
    & 0.0  & $0.154 \pm 0.055$ & $0.997 \pm 0.011$ & $16 \pm 11$ & $0.158\pm0.050$  & $1.004\pm0.011$&$0.63\pm0.46$\\
Halos 2  & 0.5  & $0.20 \pm 0.10$ & $1.006 \pm 0.016$ & $15 \pm 10$ & $0.186\pm0.085$ &$1.010\pm0.016$ &$0.52\pm0.39$\\
    & 1.0  & $0.27 \pm 0.15$ & $1.004 \pm 0.021$ & $15 \pm 10$ & $0.27\pm0.16$ & $1.009\pm0.022$&$0.58\pm0.42$\\
\hline
    & 0.0  & $0.115 \pm 0.022$ & $0.994 \pm 0.012$ & $17 \pm 12$ & $0.121\pm0.020$  &$1.003\pm0.015$ &$0.52\pm0.35$\\
Halos 3  & 0.5  & $0.21 \pm 0.11$ & $0.988 \pm 0.017$ & $14 \pm 11$ & $0.185\pm0.062$ &$0.997\pm0.018$ &$0.66\pm0.46$\\
    & 1.0  & $0.27 \pm 0.14$ & $1.000 \pm 0.025$ & $15 \pm 10$ & $0.28\pm0.15$ &$1.010\pm0.027$ &$0.78\pm0.59$\\

\noalign{\smallskip}
\hline
\hline
\end{tabular}
\\
\begin{tabular}{|c c c | c c c c c |}
\noalign{\smallskip} 
\noalign{\smallskip}
\noalign{\smallskip}
       \multicolumn{8}{c}{Redshift-space}  \\
\noalign{\smallskip}
\hline
\hline
\noalign{\smallskip}
\noalign{\smallskip} 
    &      & \multicolumn{3}{c}{Fits with $\xi^{\rm nl}_{\rm dw}$} & \multicolumn{3}{c}{Fits with $\xi^{\rm nl}_{\rm mc}$}\\
\noalign{\smallskip}
 Sample id & $z$ & $k_{\star}/(h\,Mpc^{-1})$ & $\alpha$ & $Q$&$k_{\star}/(h\,Mpc^{-1})$ & $\alpha$ &$A_{\rm mc}$\\
\noalign{\smallskip}
\hline
\noalign{\smallskip}
    & 0.0 & $0.092 \pm 0.008$ & $1.002 \pm 0.009$ & $18 \pm 12$ & $0.0969\pm0.0051$&$1.009\pm0.010$ &$0.34\pm0.21$\\
DM  & 0.5 & $0.102 \pm 0.010$ & $0.994 \pm 0.006$ & $16 \pm 11$ &$0.1074\pm0.0057$  &$1.003\pm0.008$ &$0.29\pm0.19$\\
    & 1.0 & $0.122 \pm 0.013$ & $0.993 \pm 0.005$ & $13 \pm 10$ &$0.1248\pm0.0056$ & $0.998\pm0.006$&$0.22\pm0.15$\\
\hline
         & 0.0 & $0.091 \pm 0.013$ & $0.991 \pm 0.012$ & $18 \pm 13$ & $0.096\pm0.012$ &$0.996\pm0.013$ &$0.37\pm0.26$\\
Halos 1  & 0.5 & $0.104 \pm 0.016$ & $1.000 \pm 0.012$ & $18 \pm 13$ & $0.110\pm0.012$ &$1.008\pm0.013$ &$0.52\pm0.33$\\
         & 1.0 & $0.127 \pm 0.028$ & $0.997 \pm 0.017$ & $16 \pm 12$ & $0.133\pm0.027$ &$1.004\pm0.018$ &$0.53\pm0.35$\\
\hline
         & 0.0 & $0.094 \pm 0.021$ & $0.996 \pm 0.020$ & $17 \pm 12$ & $0.099\pm0.021$ & $1.003\pm0.019$ &$0.53\pm0.39$\\
Halos 2  & 0.5 & $0.142 \pm 0.050$ & $1.001 \pm 0.020$ & $16 \pm 12$ & $0.143\pm0.046$ & $1.008\pm0.020$ &$0.57\pm0.43$\\
         & 1.0 & $0.19 \pm 0.10$   & $1.000 \pm 0.020$ & $15 \pm 11$ & $0.20\pm0.12$   & $1.000\pm0.019$ &$0.50\pm0.37$\\
\hline
         & 0.0 & $0.104 \pm 0.019$ & $0.989 \pm 0.015$ & $18 \pm 13$ & $0.109\pm0.017$ & $1.000\pm0.016$ &$0.61\pm0.41$\\
Halos 3  & 0.5 & $0.182 \pm 0.080$ & $0.982 \pm 0.017$ & $16 \pm 12$ & $0.150\pm0.045$ & $0.997\pm0.021$ &$0.93\pm0.62$\\
         & 1.0 & $0.23 \pm 0.14$   & $1.000 \pm 0.021$ & $15 \pm 11$ & $0.20\pm0.10$   & $1.016\pm0.028$ &$0.64\pm0.50$\\
\noalign{\smallskip}
\hline
\hline
\end{tabular}
\caption{
Results of applying the general fitting procedure described in 
Sec.~\ref{sec:imp} to the correlation function of the different 
samples analysed in real-space (top) and redshift-space (bottom).
The first column gives the label of the sample, as defined in 
Section 2. The second column gives the redshift output. Columns 3, 
4 and 5 show the constraints on $k_{\star}$, $\alpha$ and $Q$ obtained 
by fitting $\xi^{\rm nl}_{\rm dw}$, with the non-linear shape correction 
from Eq.~(\ref{eq:pdw_nl}). Columns 6, 7 and 8 show the constraints on 
$k_{\star}$, $\alpha$ and $A_{\rm mc}$ obtained using the model of 
Eq.~\ref{eq:xinl_rpt}. All quoted values correspond to the mean 
and 68\% c.l. of the respective parameter.
}
}\end{center}
\label{tab:params}
\end{table*}

\subsection{Redshift-space effects}
\label{sec:redshift}

When dealing with spectroscopic surveys, the distance to a galaxy 
is inferred from its measured redshift. As the redshift is affected 
by the peculiar motion of the galaxy along the line of sight, the 
radial distance so obtained does not correspond to the true distance 
to the galaxy. This changes the clustering pattern of galaxies, 
leading to differences between the power spectrum and correlation 
function measured in redshift-space, $P_{\rm s}(k)$ and $\xi_{\rm s}(r)$, 
and their true real-space counterparts. These differences are called 
redshift-space distortions. As the peculiar velocity field of galaxies 
is induced by the underlying matter distribution, it is possible to 
model the distortions that will arise in a given cosmological model.

On large scales where linear theory is applicable, the peculiar velocities 
take on the form of coherent bulk flows towards overdense regions. This leads 
to an increase in the amplitude of $P_{\rm s}(k)$ and $\xi_{\rm s}(r)$ 
compared to the real-space values. Under the assumption of linear 
perturbation theory and the plane parallel approximation, the 
scale-independent boost in power produced by this effect is given 
by \citep{kaiser87}
\begin{equation}
 S\equiv \frac{\xi_{\rm s}(r)}{\xi(r)}=\left( 1+\frac{1}{5}\beta+\frac{2}{5}\beta^2\right),
\label{eq:xi_s} 
\end{equation}
where $\beta=f/b$ and $b$ is the bias factor (which for the dark matter is simply $b=1$). 
The same factor applies for the power spectrum.

On small scales peculiar velocities are dominated by the random motions 
inside virialized structures. This makes bound structures such as dark matter 
haloes to elongated when mapped in redshift-space, an effect commonly known 
as `fingers of god'. This smearing of structure causes a damping of the 
power spectrum and the correlation function on small scales.
A complete description of redshift space distortions must include both of these regimes
\citep[for an empirical description, see][]{pd1994}.

\citet{scoccimarro04} predicted that in the case of the power spectrum, the 
boost factor of Eq.~(\ref{eq:xi_s}) is reached asymptotically on 
very large scales ($r > 100\,h^{-1}{\rm Mpc}$) and confirmed this with intermediate 
scale N-body simulations. Similar conclusions were reached with much larger 
simulations by Angulo et~al. (2008). 
For smaller scales the power is damped by random motions in a way that can be approximated 
by \citep{pd1994,park1994,cole1994}
\begin{equation}
\frac{P_{\rm s}(k)}{P(k)}=S\,\left( 1+k^2\sigma^2\right)^{-1},
\label{eq:pk_s} 
\end{equation}
where $\sigma$ is a free parameter connected to the pairwise velocity 
dispersion. This implies that in order to use information contained in 
the full shape of the power spectrum to obtain constraints on cosmological 
parameters, a detailed model of redshift space distortions must be implemented.

The open triangles of Fig.\ref{fig:all_z0}(a) show the mean two point 
correlation function measured in redshift-space from the ensemble of 
simulations, rescaled by the boost factor $S$ from Eq.~(\ref{eq:xi_s}). 
This accurately corrects the amplitude of $\xi_{\rm s}(r)$ to match 
that of the real-space $\xi(r)$ over the range of scales plotted. 
The most important effect on the correlation function of the decrease 
in power on smaller scales (as described by the extra factor 
in Eq.~(\ref{eq:pk_s})) is to produce an extra damping of the acoustic 
oscillations. This can be accurately modelled by a smaller value 
of $k_{\star}$.

Hence the treatment of redshift-space distortions in the correlation 
function is somewhat simpler than for the power spectrum, at least for 
the relatively large scales we consider here. If the amplitude of the 
correlation function is marginalized over when constraining cosmological 
parameters, the presence of the boost factor $S$ is irrelevant. 
In this case, redshift-space distortions simply produce a stronger damping 
of the acoustic oscillations, which is reflected in the fit to the 
measurements preferring smaller values of $k_{\star}$. This can be seen 
in Fig.~\ref{fig:cont_dm}, which shows the two-dimensional constraints 
on $\alpha$ and $k_{\star}$. In redshift-space, the measurements from 
the simulation prefer lower values of $k_{\star}$ than in real-space. 
The constraint on $k_{\star}$ is also tighter in redshift-space, with 
$k_{\star}=0.097 \pm 0.004\,h\,{\rm Mpc^{-1}} $ at $z=0$. This can 
also be seen in the lower panel of Fig.~\ref{fig:histo_a_k}, which 
shows the distribution of the values of $k_{\star}$ recovered from 
the $z=0$ redshift-space correlation function of the different 
realizations (dashed histogram). The best value for $k_{\star}$ in 
redshift-space is in better agreement with the prediction of 
Eq.~(\ref{eq:kstar}). The same situation is found at higher redshifts, 
where the preferred values for $k_{\star}$ are 
$k_{\star}=0.107 \pm 0.005\,h\,{\rm Mpc^{-1}}$ at $z=0.5$ and 
$k_{\star}=0.124 \pm 0.005\,h\,{\rm Mpc^{-1}}$ at $z=1$, in both cases 
lower than the corresponding values in real-space.

The constraints on the stretch factor found using redshift-space 
clustering are weaker than in real-space, with $\alpha=1.002 \pm 0.009$ 
at $z=0$, showing a considerable increase in the width of the allowed 
region for this parameter. However, this result does not show the small 
bias seen in $\alpha$ in real-space, since redshift-space distortions 
alleviate the problem of non-linear evolution at small scales. At higher 
redshift, the preferred values again towards $\alpha<1$, with 
$\alpha=0.995\pm 0.006$ at $z=0.5$ and $\alpha=0.994\pm 0.005$ at $z=1$, 
with $\alpha=1$ on the edge of the 68\% c.l. regions.

When the non-linear correction of Eq.~(\ref{eq:pdw_nl}) is applied, 
the constraint on $\alpha$ remains unaltered. As when using real-space 
measurements, the same trend towards lower values of $Q$ with increasing 
redshift is found, and the mean values for $k_{\star}$ are lower than 
the ones found without applying this non-linear correction.

Table~2 also shows the constraints obtained comparing the model of
Eq.~(\ref{eq:xinl_rpt}) against the redshift-space dark matter correlation functions.
The mean values for $k_{\star}$ recovered in this case are in excellent 
agreement with those obtained applying Eq.~(\ref{eq:xidw}).
On the other hand, the constraint on $\alpha$ show much better agreement with the
value $\alpha=1$, especially at higher redshifts. This implies that, as suggested by
\citet{crocce08}, the ansatz given by Eq.~(\ref{eq:xinl_rpt}) can also correct for the
small bias towards $\alpha < 1$ when using redshift-space information.

\begin{figure*}
\centering
\centerline{\includegraphics[width=0.85\textwidth]{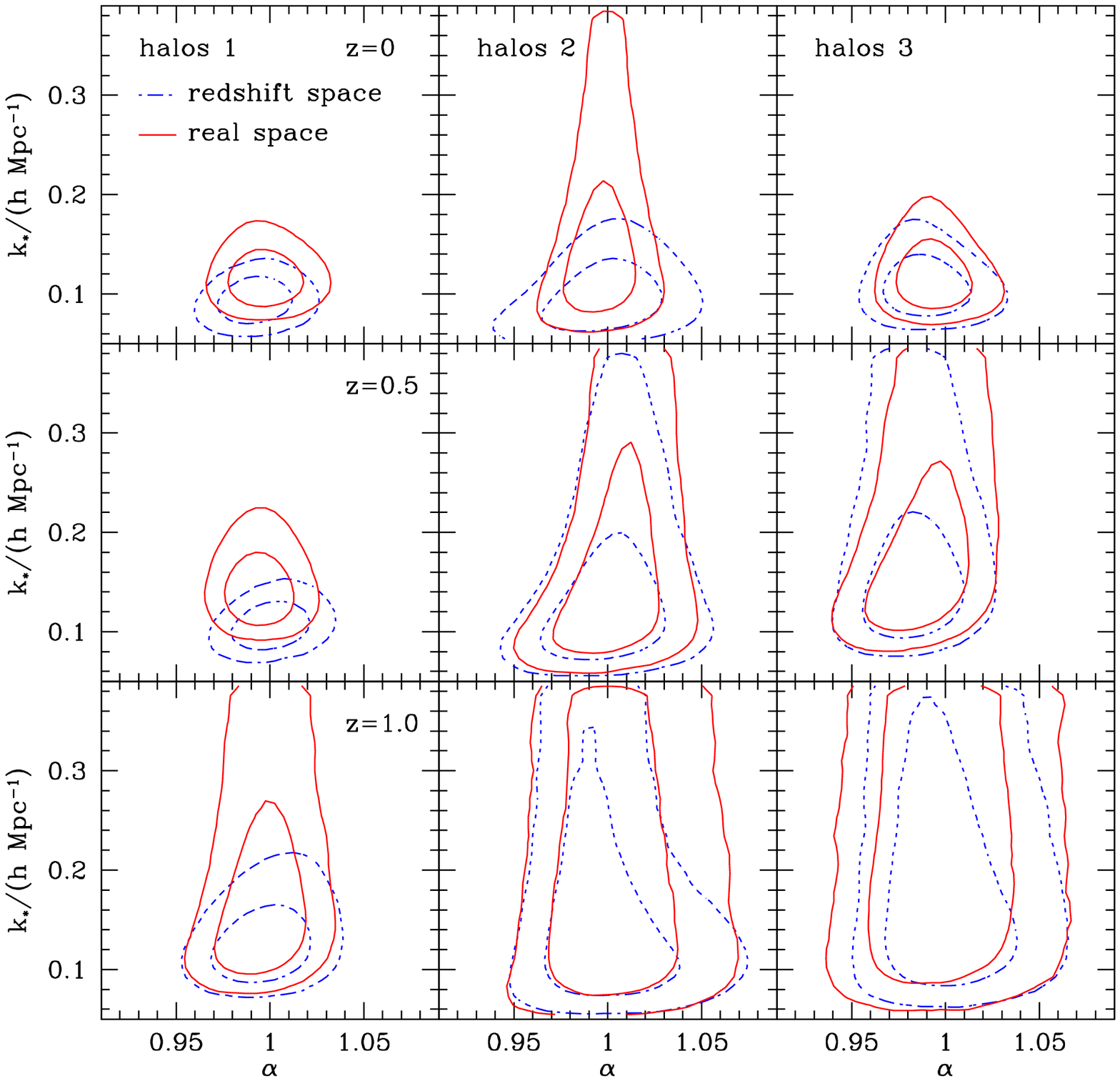}}
\caption{
Constraints on $\alpha$ and $k_{\star}$ obtained using the mean 
real-space (solid lines) and redshift-space (dashed lines) 
halo-halo correlation functions for the halo samples 1, 2 and 3 (first, middle and last
column respectively) from the ensemble of simulations at redshift $z=0$, 0.5 and 1. (upper,
middle and lower rows respectively).
}
\label{fig:cont_halos}
\end{figure*}

\subsection{Halo bias}
\label{sec:bias}

According to the current picture of galaxy formation, 
galaxies are hosted within dark matter halos. Hence, a model 
for the clustering of dark mater halos is a necessary first 
step towards an understanding of galaxy bias.
Moreover, as discused in \citet{angulo05}, galaxy clusters 
can be a useful tracer of BAO in their own right, offering an 
important cross-check of the results obtained from galaxy samples. 
Clusters are more strongly clustered than galaxies, thereby  
increasing the amplitude of the correlation function signal 
on the very scales relevant to BAO, on which the amplitude of 
the typical galaxy correlation function is very low. Future 
cluster catalogues will map in an homogeneous way volumes large 
enough for an accurate determination of the signal of the acoustic 
peak \citep{SPT, bartlett2008}. For these reasons the analysis of the 
way in which the imprints of the acoustic oscillations on the halo 
two-point correlation function change as clustering evolves is of 
great importance.

\citet{smith08} performed a detailed analysis of the halo correlation function
on the scales relevant to studies of acoustic oscillations. They focused on the
determination of the existence of shifts in the peak of the correlation function
from the predictions of linear perturbation theory. If the peak in the halo correlation
function is identified with the sound horizon at recombination in order to constraint the
dark energy equation of state parameter, these shifts introduce important biases on the
obtained constraints. As we have discused in sec.~\ref{sec:peak_pos} the position of the
peak in the correlation function should not be considered as a precise cosmological tool
since it does not correspond to the exact value of the sound horizon $s$. Instead, a detailed
modelling of the full shape of the correlation function is required.
\citet{smith08} also proposed an analytic model for the halo two point correlation function in
real-space based on solutions to the pair conservation equation using characteristic curves and
a model of the pairwise velocity distribution of the haloes. This model is able to reproduce the
main trends found in numerical simulations, showing that it contains the most relevant physical
processes that control the deviations from linear theory. Here we follow a more practical approach
and analyse the biases or systematic effects introduced in the constraints on cosmological parameters
if the modelling of sec.~\ref{sec:non-linear} is applied to describe the shape of the correlation
function of different halo samples.

On the very large scales, where halo biasing is deterministic, 
a simple linear relation is expected to hold between the correlation 
function of dark matter halos, $\xi_{\rm h}(r)$, and the linear theory 
matter correlation function, with an effective bias factor, 
$b^2_{\rm eff}=\xi_{\rm h}/\xi_{\rm lin}$, independent of scale. 
This effective bias can be computed by taking the weighted average 
of the bias $b(m)$ as a function of halo mass $m$, over the selected 
halo sample:
\begin{equation}
 b_{\rm eff}=\frac{\int_0^{\infty}\psi(m)\,b(m)n(m)\,{\rm d}m}
{\int_0^{\infty}\psi(m)\,n(m)\,{\rm d}m},
\label{eq:beff}
\end{equation}
where $n(m)$ is the mass function of dark matter halos, 
which gives the space density of halos in the mass interval 
$m$ to $m+\delta m$ \citep[e.g.][]{PS1974,bond1991,bower1991,
SMT2001,J2001}, $b(m)$ gives the bias as a function of halo mass 
$m$ \citep{mo1996a,mo1996b,jing1998,SMT2001,seljak2004} and 
$\psi(m)$ represents the mass selection function applied to 
construct the halo sample. For the cases analysed here, $\psi(m)$ 
is simply given by the limits of the mass bins, i.e. $\psi(m)=1$ if 
$m_{1} \le m \le m_{2}$ and $\psi(m)=0$ otherwise. 

The effects of halo bias on the two-point correlation function 
can be seen in Fig.~\ref{fig:all_z0}. The correlation functions of 
the different halo samples can be rescaled using a constant bias 
factor, $b^2$, to agree remarkably well with the measurement for 
the dark matter. This means that the halo bias is independent of 
scale for the pair separations plotted. For each halo sample and 
redshift analysed, we computed the bias factors $b^2$ that maximize 
the likelihood of the best fit $\xi_{\rm dw}$ model to the dark 
matter correlation function. The results are shown in Table~\ref{tab:params}. 
For $z=0$, these values are $b=1.68$, 1.84 and 2.54 for the halo samples 1, 2 
and 3 respectively. We used these bias values to rescale the halo-halo 
correlation functions in Fig.~\ref{fig:all_z0}, and to compute the 
theoretical variances for each sample as described in 
Section~\ref{sec:variance}. These values can be compared with those 
predicted by Eq.~(\ref{eq:beff}). Using the recipe for the mass function 
from \citet{J2001} and the bias function of \citet{SMT2001} 
\citep[with the modified parameters of][]{sheth2002} we find that 
Eq.~(\ref{eq:beff}) gives a good description of the bias factors with 
$b_{\rm eff}=1.62$, 1.89 and 2.53 for the same halo samples. Using the 
bias prescription from \citet{seljak2004}, the disagreement with the 
estimates from the simulations is bigger. \citet{angulo08} measured 
the bias factor as a function of halo mass directly from the {\tt BASICC} 
simulation and found good agreement with the analytic prescription we use 
here, for the modest peak heights corresponding to our halo samples.

In order to analyse any possible biases or systematic effects that can 
be introduced when using information from halo samples to constrain 
cosmological parameters, we obtained constraints on $k_{\star}$ and 
$\alpha$ by applying the model described in 
Section~\ref{sec:modelling_full_shape} to our measurements of the 
halo two-point correlation function. Our results are summarised in 
Table~\ref{tab:params1}. 
Besides the overall constant bias factors, 
which are ignored (marginalized over) by our analysis procedure, 
the estimates of $\xi_{\rm h}(r)$ show a change in the shape and 
position of the acoustic peak with respect to $\xi_{\rm lin}(r)$, 
similar to the changes seen in the dark matter correlation function. 

\begin{figure}
\includegraphics[width=0.47\textwidth]{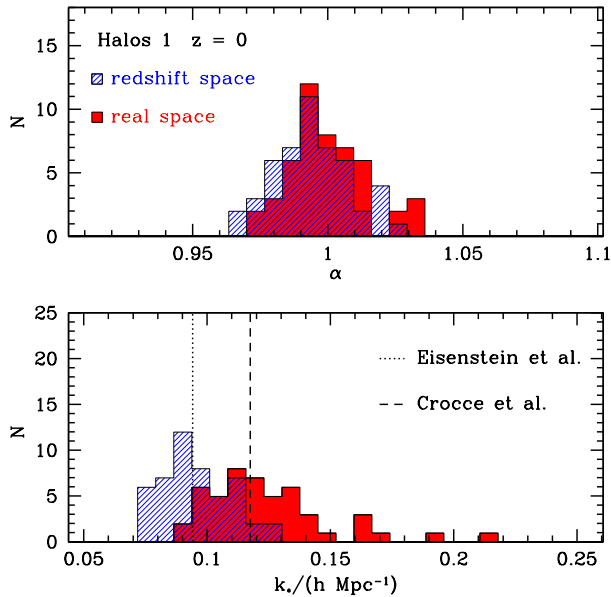}
\caption{
The distribution of the values of $\alpha$ (upper panel) and $k_{\star}$ 
(lower panel) recovered by fitting the model of Eq.~(\ref{eq:xidw}) to the $z=0$ correlation
function of halos in sample 1 measured from the different realizations in real (solid histogram) 
and redshift-space (shaded histogram). The dotted and dashed lines in the lower panel 
show the value of $k_{\star}$ predicted by Eq.~(\ref{eq:kstar}) and (\ref{eq:kstar_rpt}) respectively.
}
\label{fig:histo_a_k_m1}
\end{figure}

Fig.~\ref{fig:cont_halos} shows the two-dimensional constraints 
on $\alpha$ and $k_{\star}$ obtained using the mean halo-halo correlation
functions for halo samples 1, 2 and 3 in real (solid lines) and redshift-space
(dashed lines) measured from the ensemble of simulations. As a consequence of the
larger shot noise, the allowed regions for these parameters increases with respect
to the ones obtained using the dark matter correlation function.
The upper panel of Fig.~\ref{fig:histo_a_k_m1} shows the histogram of the values of 
$\alpha$ obtained for the $z=0$ halo-halo correlation functions for halo sample 1 
of each {\tt L-BASIIC}~{\tt II} realization, which is completely consistent with the constraints obtained from 
the mean correlation function of the ensemble. This distribution is consistent with $\alpha=1$ though it presents the same small tendency towards values of $\alpha<1$ found in \S~\ref{sec:fitting} with the dark matter.
 
\citet{smith08} performed a similar analysis, fitting the halo correlation function for different halo
samples with the linear theory $\xi(r)$ smoothed with a Gaussian filter. They found a slight tendency
towards larger shifts in the position of the acoustic peak, which implies smaller values of $k_{\star}$, with increasing mass.
We do not reproduce such a tendency here. 
The lower panel of Fig.~\ref{fig:histo_a_k_m1} shows the histogram of the values of 
$k_{\star}$ recovered from each realization. As a reference, we also show the theoretical predictions from
Eqs.~(\ref{eq:kstar}) and (\ref{eq:kstar_rpt}). Although the peak of the obtained distribution for this parameter
is consistent with the one found for the dark matter, there is a tail towards higher values of $k_{\star}$. The same
behaviour is seen in the results obtained for the remaining halo samples. This tail causes the obtained
values for $k_{\star}$ to be larger than for the dark matter, indicating that on average the halo BAO experience weaker 
damping than we find for the dark matter. This tail is more important for the higher mass bins which then creates a tendency 
towards higher mean values of $k_{\star}$ with increasing halo mass, even though the peak of the likelihood functions for this parameter are always consistent with the ones found for the dark matter.

As we saw with the dark matter, the incorporation of $Q$ 
as a free parameter of the model does not alter the constraints on 
$\alpha$. The values of $Q$ preferred by the halo measurements decrease 
at higher redshifts, when non-linear distortions are not as strong 
as at $z=0$. On the other hand, the constraints on the stretch parameter 
obtained by applying Eq.~(\ref{eq:xinl_rpt}) are completely consistent with 
$\alpha=1$ showing that this modelling is able to correct the small bias
towards $\alpha<1$ present in the previous results also in 
the case of halo samples.

The amplitude of the halo correlation function, like the matter correlation 
function, is affected by the redshift space distortions. These effects 
can be correctly described by Eq.~(\ref{eq:xi_s}) with the value of 
$S$ computed using the bias factors described above for the different 
halo samples. The values of $S$ obtained this way are shown in 
Table~2 and were used to rescale the redshift-space 
halo correlation functions plotted in Fig.~\ref{fig:all_z0}. As we found 
for the dark matter, besides this increase in the amplitude, the halo 
correlation function in redshift-space shows a stronger damping of the 
acoustic signal than in real-space. This is reflected in the allowed 
region for $k_{\star}$, which shows a preference for lower values 
than are obtained for real-space data. The constraints on $\alpha$ 
show similar behaviour to those in real-space, with some deviation 
from $\alpha=1$, which, again, is unaltered upon the inclusion of 
the non-linear correction of Eq.~(\ref{eq:pdw_nl}). As in the previous cases, 
this small bias is corrected by the implementation of the modelling of
Eq.~(\ref{eq:xinl_rpt}).

\section{Power spectrum vs correlation function}
\label{sec:pk_vs_xi}

\begin{figure}
\includegraphics[width=0.49\textwidth]{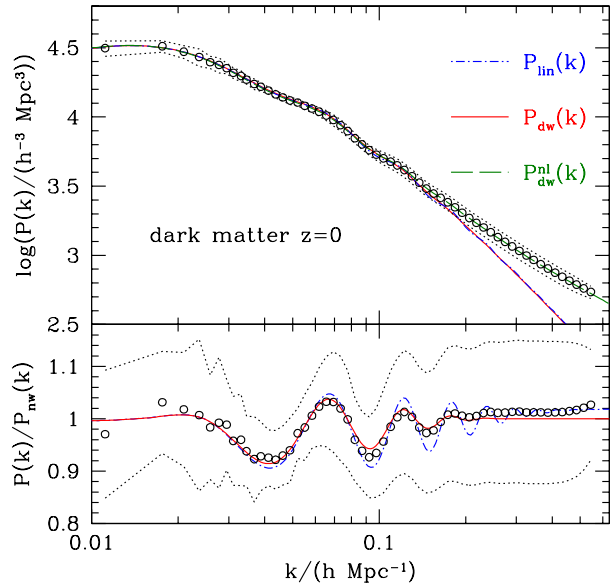}
\caption{
Upper panel: A comparison of the real-space dark matter power spectrum averaged 
over the simulation ensemble (open points) with the linear theory power spectrum 
(dot-dashed line), the `dewiggled' power spectrum from Eq.~(\ref{eq:pdw}) 
(solid line), and its non-linear version from Eq.~(\ref{eq:pdw_nl}) (dashed line) 
computed with $Q=13$. The dotted lines indicate the variance on $P(k)$ estimated 
from the ensemble. Lower panel: The ratio of these power spectra to $P_{\rm nw}(k)$. 
The results from the numerical simulations have also been divided by the 
non-linear growth factor given by Eq.~(\ref{eq:pdw_nl}).
}
\label{fig:pk_dm}
\end{figure}

The next generation of galaxy surveys will cover significantly 
larger volumes than current surveys \citep[e.g. Euclid,][or ADEPT]{SPACE}. For the 
first time, systematic effects on the scale of the BAO will become comparable 
to the sampling errors. The power spectrum and correlation function of clustering 
are affected in different ways by effects such as non-linear evolution, bias and 
redshift space distortions. In this section, we compare the constraints on the 
dark energy equation of state from these two statistics, to assess which one yields 
the smallest random and systematic errors. We compare the constraints obtained by 
fitting a model to the measured correlation function to those derived using the 
power spectrum in the same simulations as used by \citet{angulo08}.

We first recap the power spectrum and correlation function approaches to 
constraining the equation of state to highlight where differences may lie 
between them.

Various power spectrum approaches have been proposed
\citep[e.g.][]{blake03,dolney2006,wang06,percival07a,seo2007,seo2008}.
\citet{angulo08} developed the method introduced by \citet{percival07a}, 
in which no information is used from either the amplitude or large scale shape
of the power spectrum. A reference spectrum is defined from the measured power
spectrum, by applying a spline fit to the spectrum after rebinning into coarser bins.
The measured power spectrum is then divided by this reference spectrum to form a
ratio which emphasises the appearance of the BAO. The philosophy behind this
conservative approach is that dividing the measured power by a reference which is
defined directly from the measurement reduces the impact of any long wavelength
gradients in $P(k)$ induced by non-linear effects or redshift-space distortions, 
at the expense of losing some of the information contained within the power spectrum. 
Such gradients do exist in the bias factor and redshift space distortions measured 
in simulations \citep{scoccimarro04,smith07,angulo08}.
A further advantage of taking a ratio of power spectra is that this removes
the need for an accurate model of such effects. The measured ratio is then compared
to a ratio generated from a linear perturbation theory spectrum. In this case, a new
reference spectrum is generated for each linear theory $P(k)$. The linear theory
ratio is `de-wiggled' in a similar way to Eq.~(\ref{eq:pdw}). Angulo et~al. used
the wavenumber range $k=0-0.4\,h\,{\rm Mpc^{-1}}$ in their fits.

In the correlation function approach described in Section~\ref{sec:modelling_full_shape},
we model the full shape of $\xi(r)$. The key step is the damping or ``de-wiggling'' of the
BAO in the power spectrum, as described by Eq.~(\ref{eq:pdw}), before taking the Fourier transform
of the power spectrum to obtain the correlation function. 
We also apply two empirical non-linear distortions to the form of the power spectrum,
using Eqs.~(\ref{eq:pdw_nl}) and (\ref{eq:xinl_rpt}). The first approach, based on
the $Q$-model of \citet{cole05}, has little impact on the quality of the fit on the scales
used, $r=60 $--$180 \,h^{-1}\,{\rm Mpc}$. On the other hand the second model, proposed by 
\citet{crocce08} based on an ansatz inspired by RPT, gives improved constraints and  
avoids the small bias towards $\alpha < 1$ found in the original modelling.
We have demonstrated that bias and redshift space distortions 
do not change the general shape of the correlation function on these scales but can
change its amplitude. These effects are more apparent in the power spectrum approach.
The range of wavenumbers used in fitting models to the power spectrum ratio described
in the previous paragraph is wider than we use for the correlation function, which gives
a longer baseline for any gradients to become apparent. Also, when forming the correlation
function from the power spectrum, different scales are mixed together, which weakens any
gradients in $\xi(r)$. The only relevant impact of redshift space distortions for
the case of the correlation function is to increase the damping of the acoustic peak
compared with real space, which results in a lower value of the smoothing scale
$k_{\star}$ being returned by the fitting procedure. The correlation function approach
also ignores the absolute amplitude, but does use the shape of the correlation function
around the acoustic peak, as well as the shape and form of the peak itself.

We now compare how well the two approaches work in practice. As we have already seen 
in Fig.~\ref{fig:diff}, a detailed comparison of the predictions of Eq.~(\ref{eq:xidw}) 
with the mean correlation function measured from the ensemble of simulations shows small
discrepancies. These are the source of the small bias towards $\alpha<1$ found in
\S~\ref{sec:fitting}. On scales larger than the acoustic peak the model slightly
overestimates the amplitude of $\xi(r)$ which follows more closely the predictions
from linear perturbation theory. On these scales, $\xi_{\rm dw}(r)$ is almost entirely
determined by the first term of Eq.~(\ref{eq:xidw}) (Crocce \& Scoccimarro 2008; 
Smith et~al. 2008). These differences may indicate
that $G(k)$, the function that controls the damping of the harmonic oscillations,
may not be correctly described by a Gaussian on these scales. The largest differences
between the model and the results from the simulations are found on scales of
$r \lesssim 90\,h^{-1}\,{\rm Mpc}$. These discrepancies can be related to the change
in the shape of the power spectrum due to nonlinear evolution. The inclusion of the
scale-dependent correction factor of Eq.~(\ref{eq:pdw_nl}) can alleviate these differences,
but it is not enough to correct for the small bias in the constraints on $\alpha$. This
implies that the implementation of this modelling when dealing with real observational
data may lead to a slight bias in the constraints on the dark energy equation of state
$w_{\rm DE}$.
On the other hand the model based on RPT proposed by \citet{crocce08} yields tighter 
constraints and does not suffer from biases towards $\alpha < 1$.

\begin{table}
\begin{center}

\begin{tabular}{|c c c | c c |}
\noalign{\smallskip} 
\noalign{\smallskip}
\hline
\hline
\noalign{\smallskip}
\noalign{\smallskip}
 Sample id & $z$ & $\bar{b}^2$ & $k_{\star}/(h\,Mpc^{-1})$ & $\alpha$\\
\noalign{\smallskip}
\hline
\noalign{\smallskip}
    & 0.0 & 1 & $0.118 \pm 0.010$ & $1.006 \pm 0.008$    \\
DM  & 0.5 & 1 & $0.142 \pm 0.011$ & $1.002 \pm 0.007$    \\
    & 1.0 & 1 & $0.172 \pm 0.015$ & $1.000 \pm 0.006$    \\
\hline
         & 0.0 & 2.796 & $0.141 \pm 0.043$ & $0.997 \pm 0.019$    \\
Halos 1  & 0.5 & 5.601 & $0.141 \pm 0.044$ & $1.004 \pm 0.019$    \\
         & 1.0 & 12.04 & $0.132 \pm 0.039$ & $0.992 \pm 0.020$    \\
\hline
         & 0.0 & 3.323 & $0.143 \pm 0.049$ & $1.009 \pm 0.020$    \\
Halos 2  & 0.5 & 7.306 & $0.143 \pm 0.053$ & $0.972 \pm 0.020$    \\
         & 1.0 & 17.20 & $0.132 \pm 0.046$ & $0.994 \pm 0.027$    \\
\hline
         & 0.0 & 6.214 & $0.156 \pm 0.040$ & $1.002 \pm 0.015$    \\
Halos 3  & 0.5 & 13.29 & $0.157 \pm 0.046$ & $1.009 \pm 0.015$    \\
         & 1.0 & 31.61 & $0.132 \pm 0.054$ & $1.008 \pm 0.023$    \\
\noalign{\smallskip}
\hline
\hline
\end{tabular}
\caption{
The results of applying the general fitting procedure described in 
\citet{angulo08} to the real-space power spectra of the dark matter 
and the halo samples defined in sec.~\ref{sec:estimator}. The first 
column gives the label of the sample. The second column gives the 
redshift output. Column 3 gives the mean bias factor 
$\bar{b}^2=\left\langle P_{\rm hh}(k)/P_{\rm mm}(k)\right\rangle$. 
Columns 4 and 5 give the obtained constraints on $k_{\star}$ and $\alpha$. 
All quoted values correspond to the mean and 68\% c.l. on the respective 
parameter.
}
\end{center}
\label{tab:params_pk}
\end{table}

Fig.~\ref{fig:pk_dm} shows the effectiveness of the same model applied to $P(k)$. 
The upper panel shows the mean real-space dark matter power spectrum measured from 
our ensemble of simulations (open points). The dotted lines indicate the variance 
over the different realizations. Discrepancies between the linear theory power spectrum 
(dot-dashed line) and the `dewiggled' power spectrum from Eq.~(\ref{eq:pdw}) (solid line) 
are evident. Non-linear evolution distorts the shape of $P(k)$ in a way that can be 
correctly described by Eq.~(\ref{eq:pdw_nl}) (dashed line) with $Q=13$ and $A=1.5$. 
The lower panel of Fig.~\ref{fig:pk_dm} shows the ratio of these power spectra to 
$P_{\rm nw}(k)$. The results from the numerical simulations have also been divided 
by the non-linear distortion factor. Although $P^{\rm nl}_{\rm dw}$ does a good job 
of reproducing the full shape of the power spectrum and the damping of the oscillations, 
there are small discrepancies that show the limitations of the model.

In order to assess which of the two approaches is superior when used as a tool to
recover unbiased constraints from BAO measurements, we have analysed the mean
real-space power spectrum from the ensemble of simulations for the dark matter and
the same halo samples as defined in \S~\ref{sec:estimator} using the method of \citet{angulo08}.
The constraints on $\alpha$ and $k_{\star}$ obtained in this way are listed in
Table~3. \citet{angulo08} defined the mean bias factor as
$\bar{b}^2=\left\langle P_{\rm hh}(k)/P_{\rm mm}(k)\right\rangle$. The values of
$\bar{b}^2$ obtained for the different halo samples analysed are also listed in
Table~3, and are in quite close agreement with those we quote in
Table~1 for the correlation function. For the case of the dark matter,
the value of $k_{\star}$ obtained from the power sectrum shows the same tendency
to increase with redshift as was found in the analysis of $\xi(r)$. For the halo samples,
the trend of $k_{\star}$ with redshift is less evident, perhaps due to the scale
dependent bias seen in $P(k)$. There is also a tendency towards higher values of
$k_{\star}$ with increasing mass which is also suggested in the results of
\S~\ref{sec:fitting}. The precise preferred values of
$k_{\star}$ obtained in this case are in agreement with the ones obtained using the
correlation function.

The values for $\alpha$ recovered from both methods are similar, but the
allowed region for this parameter is bigger when only the information from the
oscillations in $P(k)$ is used. The extra information from the shape of the correlation
function helps to improve the constraints on $\alpha$ resulting in a smaller allowed
region for this parameter. This test implies that the simple model for the correlation 
function described in \S~\ref{sec:non-linear} can perform better than the methods 
in which the scale of the acoustic oscillations is extracted from the power spectrum. 
In a more realistic situation, when other cosmological parameters 
are included in the analysis, the extra
information contained in the shape of the correlation function would make this approach
seem even more attractive, since it can also be used to improve constraints on
parameters, such as $\Omega_{\rm m}$. In the analysis of CMB data, this parameter shows
a strong degeneracy with $w_{\rm DE}$, and therefore a better constraint on the matter
density would also result in tighter bounds on the dark energy equation of state.

Finally, the mean values of the stretch parameter recovered from the dark matter power spectrum show a small bias towards 
$\alpha>1$ which decrease with increasing redshifts. \citet{crocce08}, estimated the mode-coupling
shifts that would affect the constraint on $\alpha$ using the method of \citet{angulo08}. Their predictions are
in good agreement with our results, suggesting that this bias might be attributed to the systematic effects
introduced by the mode-coupling power spectrum $P_{\rm mc}(k)$.
This deviation from $\alpha=1$ is smaller than the sampling variance for 
the volume of the individual {\tt L-BASICC}\,{\tt II} realizations. 
Future surveys like Euclid or ADEPT will cover volumes that
are close to the total volume simulated by our ensemble ($V=120\,{\rm Gpc}^3h^{-3}$).
Rescaling the obtained covariance matrix by a factor $1/50$ (which then corresponds to the covariance of the mean correlation function from our ensemble) we can gauge 
the effect of this bias for these surveys. In this case, the error associated 
with the value of $\alpha$ at $z=0$ will be $0.008/\sqrt{50}\approx0.001$, which then 
implies a 5 sigma detection of the mode-coupling shifts for the dark matter.
Instead, the results obtained from the correlation function applying the 
modelling of Eq.~(\ref{eq:xinl_rpt}) are more consistent with $\alpha=1$, 
even after re-scaling the error to the full volume of our ensemble. This 
shows that with this modelling, the correlation function is more likely 
to produce unbiased constraints that the power spectrum.
In principle, the simple ansatz of Eq.~(\ref{eq:xinl_rpt}) can be extended to the analysis of the power spectrum. Nevertheless, due to the fact that in Fourier 
space, redshift space distortions and halo bias do show important scale dependence, it will be necessary to follow an approach similar to \citet{angulo08}, dividing the measured power spectrum by a smooth function. This sacrifice of information will always have consequences for the obtained constraints.

\section{Conclusions}
\label{sec:conclusions}

In this paper we have addressed two main questions: 1) What is the relation between 
the ``acoustic peak" in the two-point correlation function and the sound horizon scale? 
and 2) Which is the better statistic to use to constrain the dark energy equation of 
state, the correlation function or the power spectrum? Future galaxy surveys will cover 
volumes one or more orders of magnitude larger than existing surveys. For the first 
time, systematic effects in the appearance of the BAO and in their interpretation will 
be comparable to the random or sampling variance errors on the measurement of the 
two point statistics. It is therefore essential to understand these systematics and to 
improve theoretical modelling of the BAO so that they can be used to provide optimal 
and unbiased estimates of the values of cosmological parameters. 

It is a common misconception that the location of the acoustic peak in the correlation 
function is identical to the size of the sound horizon at recombination. 
Indeed, several 
recent studies \citep{guzik2007,smith08,crocce08} have focused on showing that the acoustic 
peak in the correlation function is shifted and distorted relative to the prediction 
of linear perturbation theory.
What has not been widely appreciated before is that
even in linear perturbation theory, the centroid or maximum of the peak in $\xi(r)$
does {\it not} coincide with the sound horizon scale at 
the level of accuracy required to fully exploit the constraints that can be derived 
from forthcoming galaxy surveys. The size of the error in the distance scale which would 
result from making this incorrect assumption is of the order of 1\%, which is already 
close to the random error in the distance scale forecast for ongoing surveys.

In order to use the correlation function on large scales to constrain the values 
of the cosmological parameters, it is therefore necessary to model the form of the 
correlation function. By using a physically motivated model, a natural and consistent 
connection is made between the cosmological parameters and the form of the 
correlation function. There are several effects which have to be taken into 
consideration to produce a complete model: the non-linear growth of fluctuations, 
the possible scale dependence of the bias between the chosen tracer of the density 
field and the underlying matter fluctuations and the distortions introduced by using 
distance measurements which are inferred from redshifts, commonly referred to as 
redshift-space distortions.  In order to assess the impact of these effects on the correlation 
function, we used an ensemble of 50 very large volume N-body simulations, the {\tt L-BASICC}\,{\tt II}. 
These simulations are analogous to the {\tt L-BASICC} ensemble used by \citet{angulo08} 
for a similar analysis of the acoustic oscillations in the power spectrum. We measured 
the correlation function of the dark matter and of different halo samples in both 
real and redshift-space in each of the realizations in the ensemble. 

It turns out that bias, general non-linear evolution and redshift-space distortions 
are much simpler to deal with in the case of the correlation function than they are 
for the power spectrum. On the scales we consider ($60 < (r/h^{-1}{\rm Mpc})< 180$), 
these effects primarily result in a change in the amplitude of the measured correlation 
function and do not alter its shape. By ``general'' non-linear evolution, we mean the 
change in the overall ``coarse grained'' shape of the power spectrum. The crucial 
effect to include in the model is the damping of the higher order harmonic oscillations 
due to non-linear evolution, as given by Eq.~(\ref{eq:pdw}). Whilst we find that a 
simple damping of the linear perturbation theory power spectrum is able to reproduce 
the correlation function we measure from the simulations, some discrepancies remain 
between the model and the simulation results on scales both larger and smaller than 
the acoustic peak. These differences lead to slightly biased constraints on the stretch 
factor $\alpha$, which in turn will yield small biases in the estimates of the dark 
energy equation of state parameter $w_{\rm DE}$. The model is only marginally improved 
on incorporating an empirical model for non-linear distortions (which could also be used 
to describe a scale-dependent bias). This $Q$-model, introduced by Cole et~al. (2005), 
is widely used in the literature but needs to be refined to extract the full information 
from the correlation function. 
On the other hand, the implementation of the model of 
Eq.~(\ref{eq:xinl_rpt}) helps to bring the constraints on the stretch parameter in closer
agreement with the value $\alpha=1$. This parametrization was proposed by \citet{crocce08}
and is based on RPT. 

To address the question of which two-point statistic is the more powerful for extracting 
the BAO, we repeated the power spectrum analysis carried out by Angulo et~al. for several 
of the samples we used to measure correlation function. The analysis of Angulo et~al. can 
be viewed as a conservative approach \citep[see also][]{percival07a,percival07c}. The measured 
power spectrum is divided by a ``wiggle-free'' reference spectrum which is defined using 
the measured spectrum itself, without any modelling. No information is used about the 
amplitude of the spectrum or its overall shape. By taking a ratio in this way, the BAO 
are isolated and the impact of scale dependent effects in the power spectrum is minimized. 
Similar ratios generated using linear perturbation theory, and applying some 
``de-wiggling" or damping of the BAO are compared to the measured power ratio. We compared 
the constraints obtained from the model for $\xi(r)$ described in Sec.~\ref{sec:non-linear} 
and the ones obtained by applying the fitting procedure of \citet{angulo08}. This 
simple test shows that a method that uses the full large-scale shape of the correlation 
function can provide tighter constraints on $w_{\rm DE}$ than one in which the scale of 
the acoustic oscillations is extracted from the power spectrum. The extra information  
contained in the shape of the correlation function can also be extremely useful to constrain 
other cosmological parameters, such as $\Omega_{\rm m}$. In the analysis of CMB data, 
$\Omega_{\rm m}$ shows a strong degeneracy with $w_{\rm DE}$, and so the use of the 
correlation function would result in tighter bounds on the dark energy equation of state.

The constraints on the stretch parameter obtained with the power spectrum show a small bias
towards $\alpha >1$ which is consistent with the predictions of \citet{crocce08} based on 
the effect of the mode-coupling power spectrum. This bias will become much more important
for future galaxy surveys which will have much smaller random errors. 
Instead, the constraints obtained using Eq.~(\ref{eq:xinl_rpt}) to model the
correlation function are much more consistent with $\alpha=1$ even for the large volumes
that will be covered by these surveys.

The ability of Eq.~(\ref{eq:xinl_rpt}) to improve the constraints obtained from the correlation
function suggests that the recent advances in the modelling of the acoustic oscillations using renormalized
perturbation theory \citep{crocce08} may provide a basis for a more
accurate theoretical model of the full shape of the correlation function
and power spectrum.
Nonetheless the problem is aggravated by the need to model
scale dependent galaxy bias, which may have already become the strongest limitation
on the use of large scale structure information to obtain constraints on cosmological
parameters \citep{sanchez08}. An accurate model capable of describing all sources of
distortions must be implemented if the determinations of the galaxy correlation function and
power spectrum from future surveys are to achieve competitive constraints on the
dark energy equation of state. It is clear that realistic numerical simulations which
are able to model the growth of structure in the dark matter and which can incorporate
a physical model for galaxy formation will be essential to guide the development of
such models. The use of both of these statistics will give the complementary information
required to improve the model of Eq.~(\ref{eq:pdw}), and will hopefully allow the full
potential of BAO measurements as probes of the nature of dark energy to be reached.

\section*{Acknowledgments}

We would like to thank Mart\'in Crocce, Hee-Jong Seo and Takahiko Matsubara
for useful comments and discussions. AGS acknowledges the hospitality of the
Department of Physics at the University of Durham where part of this work was
carried out. CMB is funded by a Royal Society University Research Fellowship; 
RA acknowledges receipt of a Dorothy Hodgkin PhD scholarship.
This work was supported by STFC, the European Commission's ALFA-II programme 
through its funding of the Latin-american European Network for Astrophysics 
and Cosmology (LENAC), the Consejo Nacional de Investigaciones
Cient\'{\i}ficas y T\'ecnicas (CONICET) and the Royal Society, through the award of an 
International Incoming Short Visit grant.





\end{document}